\documentclass[10pt]{article}

\usepackage{amsmath,graphicx,bbm,amssymb}
\usepackage{caption}
\usepackage{color}
\usepackage{tikz}
\usetikzlibrary{shapes.multipart}
\usetikzlibrary{shapes.misc, positioning}
\usepackage{fancyhdr}
\usepackage{colortbl}
\usepackage{marvosym}
\usepackage[hmargin=1.5cm, top=2cm, bottom=2cm]{geometry}
\usepackage[colorlinks=true,linkcolor=blue,citecolor=blue,urlcolor=blue]{hyperref}
\usepackage{natbib}
\usepackage[export]{adjustbox}
\usetikzlibrary{arrows}

\newcommand{\module}[1]{\left\vert #1 \right\vert}
\newcommand{\norme}[1]{\left\vert\left\vert #1 \right\vert\right\vert}
\newcommand{\para}[1]{\left(#1\right)}

\newcommand{\aver}[1]{\left\langle #1 \right\rangle}

\newcommand{\xth}[1]{#1^{\text{th}}}

\newcommand{\rz}{r_0}

\newcommand{\cnh}{C_n^2(h)}
\newcommand{\cnhl}{C_n^2(h_l)}

\newcommand{\rbb}{\boldsymbol{r}}
\newcommand{\rbun}{\boldsymbol{r}_1}
\newcommand{\rbdeux}{\boldsymbol{r}_2}
\newcommand{\rhob}{\boldsymbol{\rho}}

\newcommand{\TTout}{\boldsymbol{\text{P}}_\text{TTR}}
\newcommand{\TTonly}{\boldsymbol{\text{P}}_\text{TT}}
\newcommand{\otf}[1]{\text{OTF}_{#1}}

\newcommand{\atf}[1]{\text{ATF}_{#1}}
\newcommand{\cov}[1]{\mathcal{C}_{#1}}

\newcommand{\Hdm}{\mathcal{P}_\text{DMR}}

\newcommand{\pas}{\phi_\text{s}}
\newcommand{\pal}{\phi_\text{l}}
\newcommand{\pan}{\phi_\text{n}}
\newcolumntype{P}[1]{>{\centering\arraybackslash}p{#1}}
\newcommand{\thetavec}{{\boldsymbol{\theta}}}

\newcommand{\rhovec}{\boldsymbol{\rho}}

\newcommand{\rvec}{{\mathbf{r}}} 

\usepackage{titling}
\usepackage[affil-it]{authblk}

\title{\LARGE Off-axis point spread function characterisation in laser-guide star adaptive optics systems} 

\author{
	O. Beltramo-Martin,$^{1}$\thanks{E-mail: olivier.beltramo-martin@lam.fr}
	C.M. Correia,$^{1}$
	E. Mieda,$^{2}$
	B. Neichel,$^{1}$
	T. Fusco,$^{1,3}$\\
	G. Witzel,$^{4}$
	J. Lu,$^{5}$
	J.-P. V\'eran$^{6}$
}
\affil{
	\small $^{1}$Aix Marseille Univ., CNRS, LAM, Laboratoire d'Astrophysique de Marseille, 38 rue F. Joliot-Curie, 13388 Marseille, France\\
	$^{2}$Subaru Telescope, 650 North A'ohoku Place, Hilo, HI, 96720, USA\\
	$^{3}$ONERA, BP. 72, F-92322 Chatillon Cedex, France\\
	$^{4}$Department of Physics and Astronomy, UCLA, Los Angeles, CA 90095-1547, United States\\			
	$^{5}$Department of Astronomy, UCB, Berkeley, CA 94720-3411, United States\\
	$^{6}$National Research Council – Herzberg, 5071 W. Saanich Rd., Victoria, BC, V8V 2R5, Canada
}
\date{}

\begin{document}
	
\maketitle

\begin{abstract}
Adaptive optics~(AO) restore the angular resolution of ground-based telescopes, but at the cost of delivering a time- and space-varying point spread function~(PSF) with a complex shape. PSF knowledge is crucial for breaking existing limits on the measured accuracy of photometry and astrometry in science observations. In this paper, we concentrate our analyses on anisoplanatism signature only onto PSF: for large-field observations~(20") with single-conjugated AO, PSFs are strongly elongated due to anisoplanatism that manifests itself as three different terms for Laser-guide star~(LGS) systems: angular, focal and tilt. We propose a generalized model that relies on a point-wise decomposition of the phase and encompasses the non-stationarity of LGS systems. We demonstrate it is more accurate and less computationally demanding than existing models: it agrees with end-to-end physical-optics simulations to within 0.1\% of PSF measurables, such as Strehl-ratio, FWHM and fraction of variance unexplained. Secondly, we study off-axis PSF modelling is with respect to $\cnh$ profile (heights and fractional weights). For 10~m class telescope, PSF morphology is estimated at 1\%-level as long as we model the atmosphere with at least 7 layers whose heights and weights are known respectively with 200m and 10\%-precision.
As a verification test we used the Canada's NRC-Herzberg HeNOS testbed data, featuring four lasers. We highlight capability of retrieving off-axis PSF characteristics within 10\% of fraction of variance unexplained, which complies with the expected range from the sensitivity analysis. Our new off-axis PSF modelling method lays the ground-work for testing on-sky in the near future.
\end{abstract}

\section{Introduction}
Ultimate scientific exploitation of astronomical observations is made recurring to post-processing techniques - commonly called data-reduction. The current way to process scientific images relies commonly on standard packages, such as StarFinder~\citep{Diolaiti2000}, SeXTRACTOR~\citep{Bertin1996}, or DAOPHOT~\citep{Stetson1987}. Nonetheless, past studies~\citep{Fritz2010,Yelda2010} have quantified an astrometry error breakdown on the Galactic Center~(GC), that pointed out the Point spread function~(PSF) model accounts at 50-60~\% level on the astrometry accuracy.
Photometry measurements are impacted by PSF mis-knowledge as well~\citep{Shodel2010,Sheehy2006}, but a recent analysis~\citep{Ascenso2015} has unveiled that photometry measurements accuracy is improved from 0.2 down to 0.02 mag by providing StarFinder with the exact PSF.

Besides, ground-based astronomy is improved thanks to Adaptive optics~(AO) that allows to restore the angular resolution to nearly the diffraction limit. However AO produces a PSF that varies across space and time and does not match standard parametric models, such as Gaussian or Moffat functions. Retrieving the AO PSF shape from focal-plane observations might be also compromised when observing crowded fields suffering from large source confusion~\citep{Turri2017,Shodel2010,Lu2013,Ghez2008}, or deep cosmological fields where no PSF is observed~\citep{Falomo2008,Schramm2013}, calling for alternative approach as PSF reconstruction~(PSF-R)~\citep{Veran1997,Gilles2012}.
PSF-R is a data processing approach that delivers the PSF from AO control loop data, without any priors on its shape. PSF-R reliability has been demonstrated on-sky multiple times~\citep{Veran1997,Flicker2008,Jolissaint2015,Martin2016JATIS}. PSF is reconstructed as a convolution of independent patterns that characterize different physical limitations of AO systems~\citep{Veran1997,Gilles2012}. In this paper, we focus on the specific step of PSF extrapolation that relies on anisoplanatism~\citep{Fried1982} and accounts for PSF spatial variations across the field. The AO system measures the incoming distorted wave-front in the particular direction of the guide star; wave-fronts that propagate along another direction do not cross the exact same turbulence are partially compensated. This introduces an anisoplanatism error that grows with the angular separation from the guide star and the seeing of altitude layers. 
AO systems can also guide on Laser guide stars~(LGS)~\citep{Foy1985} which are focused in a range from 80 up to 100~km. When the LGS beam propagates as a spherical wave downwards to the pupil, it crosses only a portion of the turbulence above the telescope. The phase aberrations in the science direction are not fully corrected; it includes an additional term in the residual phase that stands as the cone effect or focal anisoplanatism.

On top of that, probing the wave-fronts using LGSs does not grant access to the wave-front angle of arrival because of the round trip of the light from telescope to the sodium (Na) layer. The PSF position in the focal plane is stabilized by measuring the wave-front angle of arrival using an Natural guide star~(NGS) whose location may be different from the scientific target. This angular separation yields an anisoplanatism effect on tip-tilt modes only, named as tip-tilt anisoplanatism or anisokinetism~\citep{Winick1988,Fried1996,Ellerbroek2001,Flicker2003,Correia2011}. 

Reconstructing the off-axis PSF for LGS-based systems requires a model of the angular, focal and tilt anisoplanatism plus the relevant input parameters the model depends on which are the vertical distribution of turbulence known as the $\cnh$ profile. Such models exist in the literature, but are either not laser-compliant~\citep{Tyler1994,Rigaut1998,Jolissaint2010}, or not numerically efficient~\citep{Fusco2000,Britton2006,Flicker2008}, particularly for future AO systems on next 40~m class telescopes with large number of degrees of freedom. The $\cnh$ profile is accessible either from internal methods handling AO telemetry of multiple guide-stars~\citep{Ono2017,Martin2016L3S,Guesalaga2016}, or from external profilers~\citep{Wilson2002,Tokovinin2007,Osborn2013,Osborn2015} or more recently from meso-scale models~\citep{Masciadri2017}.

For single guide-star AO systems, the $\cnh$ is not measurable from AO telemetry, calling for external profilers for modelling the off-axis PSF. However, studies~\citep{Ono2017} have highlighted discrepancies on the retrieved profile by comparing outputs of external to internal profiling techniques. Consequently, using an external $\cnh$ profile along a likely different line-of-sight from the observations might degrade the accuracy of PSF modeling.
Besides, predicting performance of future AO-based instrument relies on a $\cnh$ profile as well; we must constrain the $\cnh$ accuracy required for retrieving faithfully AO-corrected PSFs.

We propose in this paper an investigation of how the PSF is degraded regarding the accuracy on the $\cnh$. Yet, such an analysis is compromised if relies on on-sky observations, in a way the real $\cnh$ is not be perfectly known; downstream results would be contaminated by anisoplanatism model errors, such as the discrete number of layers or the exact LGS height that is not perfectly identified for instance.
As an alternative we have applied this approach on the HeNOS Multi-conjugated AO~(MCAO) testbed designed to be the demonstrator for NFIRAOS~\citep{Conan2010}. We lay in this paper all the theoretical background our near-future work will be based on, that will be particularly focused on PSF characterisation application to crowded fields observation, such as the Galactic center.

This paper is organized as follows. In Sect.~\ref{S:PSFR}, we derive a generalized anisoplanatism model that combines both focal and tip-tilt anisoplanatism recurring to an accurate and fast implementation. In Sect.~\ref{S:comparison} it is compared to physical-optics simulations on a 10~m telescope. We consider morphological PSF metrics, as Strehl ratio, FWHM and fraction of variance unexplained, as science metrics~(photometry and astrometry for tight binaries) as well. In Sect.~\ref{S:sensitivity}, we present how those criteria are sensitive to $\cnh$ accuracy, including number of layers and weight/height precision. We validate the meaning of this approach using observations on the HeNOS bench in Sect.~\ref{S:HENOS}. We conclude in Sect.~\ref{S:conclusions}.

\section{Spatial PSF extrapolation}
\label{S:PSFR}

\subsection{Anisoplanatism transfer function}

We define $\phi_\varepsilon(\rvec,t)$ as the residual wave-front delivered by the AO system. At the focal plane downstream the AO system, the long-exposure Optical Transfer Function~(OTF) in the science direction 1 is~\citep{Roddier1981}
\begin{equation}
\begin{aligned}	
	\otf{1}(\rhovec/\lambda) =\dfrac{1}{S} \iint_{\mathcal{\mathcal{P}}} \mathcal{P}(\rbb)\mathcal{P}(\rbb+\rhovec)\cdot
	\exp\para{-\dfrac{1}{2}D_{\phi_\varepsilon}(\rbb,\rhovec)}\boldsymbol{dr},
\end{aligned}
\label{E:otfdef}
\end{equation}
where $\mathcal{P}$ is the pupil function, $\rbb$ and $\rhob$ the location and separation vectors in the pupil and $S$ is the area of the telescope aperture which normalizes the  PSF to unit energy and $D_{\phi_\varepsilon}(\rbb,\rhovec)$ is the residual phase structure function defined by
\begin{equation} 
\label{E:dphidef}
	D_{\phi_\varepsilon}(\rbb,\rhovec) = \aver{\module{\phi_\varepsilon(\rbb,t) - \phi_\varepsilon(\rbb+\rhovec,t)}^2}_t,
\end{equation}
where $\aver{x(t)}_t$ denotes the temporal average of process $x$. Under the assumption that the anisoplanatism term is not correlated to other AO error terms, such as servo-lag or aliasing, $D_{\phi_\varepsilon}(\rbb,\rhovec)$ can be split as follows
\begin{equation}
	D_{\phi_\varepsilon}(\rbb,\rhovec) = D_{0}(\rbb,\rhovec) + D_{\Delta}(\rbb,\rhovec),
	\label{E:dphisum}
\end{equation}
where $D_{0}(\rbb,\rhovec)$ characterizes the AO residual phase structure function in the AO guide star direction~(NGS or LGS as well), while $D_{\Delta}(\rbb,\rhovec)$ is the structure function of the anisoplanatic phase $\phi_\Delta(\rbb,t)$ defined as
\begin{equation} 
\phi_\Delta(\rbb,t) = \phi_1(\rbb,t) - \phi_0(\rbb,t),
\label{E:phiDelta}
\end{equation}
where $\phi_1$ and $\phi_0$ refer to the atmospheric phase in direction 1~(science) and direction 0~(guide star). As explicitly mentioned in Eq.~\ref{E:otfdef}, the phase structure is a function of both position and separation in the pupil. For on-axis PSF reconstruction, we handle the residual phase as a stationary process~\citep{Veran1997}; we assume $D_{0}(\rbb,\rhovec)$ is a function of separation only, that allows to average $D_{0}$ over the pupil location as
\begin{equation}
\bar{D}_{\text{0}}(\rhovec) = \dfrac{\iint_{\mathcal{P}} \mathcal{P}(\rbb)\mathcal{P}(\rbb+\rhovec) D_{0}(\rbb,\rhob) \boldsymbol{dr}}{\iint_{\mathcal{P}}\mathcal{P}(\rbb)\mathcal{P}(\rbb+\rhovec)\boldsymbol{dr} },
\end{equation}
that allows to pull the exponential term in Eq.~\ref{E:otfdef} out of the integral, which makes for an easier and more convenient numerical implementation.\\

For the LGS case, \citep{Flicker2008} has pointed out the stationarity hypothesis is no longer accurate and degrades the PSF model because the cone effect. We have confirmed on our side that this assumption degrades PSF metrics~(introduced in Sect.~\ref{S:metrics}) at the level of 5\%, while it is maintained to less than 1\% in the NGS case. Because we have the numerical ability to derive the full calculation $D_{\Delta}(\rbb,\rhovec)$, the formalism we present here does not rely on the stationarity hypothesis. We include into $D_0$ all AO-residual in the guide star direction and separate any focal and angular anisoplanatism effect into $D_{\Delta}$, in a manner we can still apply the pupil-averaged process on $D_0$.

The OTF reduces to the expression
\begin{equation}
\begin{aligned}
\otf{1}(\rhovec/\lambda) =\dfrac{1}{S} \exp\para{-\dfrac{1}{2}\bar{D}_{0}(\rhovec)}\cdot
\iint_{\mathcal{P}}  \mathcal{P}(\rbb)\mathcal{P}(\rbb+\rhovec)\exp\para{-\dfrac{1}{2}D_{\Delta}(\rbb,\rhovec)} \boldsymbol{dr},
\end{aligned}
\label{E:otfsci}
\end{equation}

We introduce $\otf{\text{DL}}(\rhovec/\lambda)$ as the diffraction-limit OTF that characterizes the angular frequencies distribution imposed by the pupil shape
\begin{equation}
	\otf{\text{DL}}(\rhovec/\lambda) = 1/S\iint_{\mathcal{P}}  \mathcal{P}(\rbb)\mathcal{P}(\rbb+\rhovec)\boldsymbol{dr}
	\label{E:otfDL}
\end{equation}
that allows to write equation~\ref{E:otfsci} as
\begin{equation} 
\otf{1}(\rhovec/\lambda) = \otf{0}(\rhovec/\lambda) \cdot \atf{\Delta}(\rhovec/\lambda),
\label{E:gs2sci}
\end{equation}
where
\begin{equation} 
\otf{0}(\rhovec/\lambda) = \otf{\text{DL}}(\rhovec/\lambda)\cdot\exp\para{-\dfrac{1}{2}\bar{D}_{0}(\rhovec)},
\label{E:otf0}
\end{equation}
and
\begin{equation} 
\atf{\Delta}(\rhovec/\lambda) = \dfrac{\iint_{\mathcal{P}}  \mathcal{P}(\rbb)\mathcal{P}(\rbb+\rhovec)\cdot\exp\para{-\dfrac{1}{2}D_{\Delta}(\rbb,\rhovec)}\boldsymbol{dr} }{\otf{\text{DL}}(\rhovec/\lambda)}
\label{E:atf}
\end{equation}
is the Anisoplanatism Transfer Function~(ATF) as introduced by~\citep{Fusco2000} which is a convenient formulation to derive the residual OTF anywhere in the field from the knowledge of the on-axis OTF. A practical result to extrapolate the PSF anywhere in the field consists of computing
$\otf{2}$. From Eqs.~\ref{E:gs2sci} and~\ref{E:atf} comes
\begin{equation}
	\otf{2}(\rhovec/\lambda) = \otf{1}(\rhovec/\lambda)  \cdot \dfrac{\otf{\Delta_2}(\rhovec/\lambda)}{\otf{\Delta_1}(\rhovec/\lambda)},
	\label{E:otf2t1}
\end{equation}
which introduces an "OTF ratio" that is the key quantity for deriving the OTF from an direction to any other. The PSF is seamlessly computed from the Fourier transform of the OTF.

In summary, the capability to model a PSF anywhere in a field requires, firstly to have access to a  PSF in a direction 1 using either sources extraction method, parametric modelling or PSF-reconstruction, secondly be capable of computing accurately an OTF ratio. This latter is the main focus of following sections.

\subsection{Modelling anisoplanatism}

We now focus on the implementation of the phase structure function $D_{\Delta}(\rbb,\rhovec)$ for computing the ATF in Eq.~\ref{E:atf}. Results in the remainder of this paper rely on the calculation of the anisoplanatic covariance $\mathcal{C}_\Delta(\rbb,\rhovec)$ which writes
\begin{equation} 
\mathcal{C}_\Delta(\rbb,\rhovec) = \aver{\phi_\Delta(\rbb,t)\phi^t_\Delta(\rbb+\rhovec,t)}
\label{E:cphi0}
\end{equation}
from which $D_\Delta(\rbb,\rhovec)$ is
\begin{equation}
D_\Delta(\rbb,\rhovec) = 2\times\para{\mathcal{C}_\Delta(\rvec,0) - \mathcal{C}_\Delta(\rbb,\rhovec)}.
\end{equation}
The problem of modeling anisoplanatism is now a matter of describing how the atmospheric phase is spatially correlated. In general terms, phase is described as a linear combination of  $n_m$ (orthonormal) $\mathcal{M}$ modes, yielding
\begin{equation}
\begin{aligned}
	&\phi_1(\rbb,t) = \sum_{i=1}^{n_m} a_i(t)\times\mathcal{M}_i(\rbb)\\
	&\phi_0(\rbb,t) = \sum_{i=1}^{n_m} b_i(t)\times\mathcal{M}_i(\rbb),	
\end{aligned}
\end{equation}
where $\boldsymbol{a}(t)$ and $\boldsymbol{b}(t)$ are respectively modal coefficients in the science and guide star directions. The anisoplanatic phase $\phi_\Delta(\rbb,t)$ becomes
\begin{equation} \label{E:phiDelta_modal}
\phi_\Delta(\rbb,t) = \sum_{i=1}^{n_m} \Delta_i(t)\times\mathcal{M}_i(\rbb),
\end{equation}
with $\Delta_i(t) = (a_i(t) - b_i(t))$.

Combining Eqs.~\ref{E:cphi0} and~\ref{E:phiDelta_modal} gives the following expression
\begin{equation} 
\begin{aligned}
\cov{\Delta}(\rbb,\rhovec) = \sum_{i=1,j}^{n_m} \aver{\Delta_i\Delta_j^t}\times\mathcal{M}_i(\rbb)\mathcal{M}^t_j(\rbb + \rhovec),
\end{aligned}
\label{E:covDelta_modal}
\end{equation}
where 
\begin{equation}
	\Delta_i\Delta_j^t = {a_i(t)a^t_j(t)} + {b_i(t)b^t_j(t)} - {a_i(t)b^t_j(t)} - b_j(t)a^t_i(t).
\end{equation}
Eq.~\ref{E:covDelta_modal} reminds the $U_{ij}$ formalism introduced by~\citep{Veran1997} for the computation of the residual phase structure function; it simplifies to the multiplication $\mathcal{M}_i(\rbb)\mathcal{M}_j(\rbb + \rhovec)$ when deriving the covariance.
On top of that, anisoplanatism modeling requires the calculation of modal coefficients correlation $\aver{\Delta_i\Delta_j^t}$. \citep{Fusco2000} has derived the ATF using a Zernike expansion of the phase, based on Zernike coefficients spatial correlation provided in~\citep{Chassat1989}. Although this study has delivered successful results on a 8~m telescope, its application on a 40~m class telescope is computationally-demanding because of the $ \mathcal{M}_i(\rbb)\mathcal{M}_j(\rbb + \rhovec)$ derivations. An alternative computation approach was proposed by~\citep{Gendron2006} to tackle the numerical complexity of the $U_{ij}$ technique. However, although focal anisoplanatism could be covered in using covariance terms introduced by~\citep{Molodij1997}, it would not be as accurate as the NGS case because the intrinsic stationarity assumption of the Zernike expansion.

To decrease the numerical complexity whilst maintaining model accuracy, we may resort to a spatial frequency basis~\citep{Rigaut1998,Jolissaint2010}. Although such a technique is efficient from the numerical computation point of view, statistical independence of  Fourier modes assume underlying stationarity; the long-exposure OTF is derived from the AO residual phase Power Spectrum Density~(PSD), although such a description is only accurate for linear and space-invariant systems. In order to generalise it to the LGS case, correlations between the residual phase errors at different frequencies must be included~\citep{VanDam2006,Flicker2008} due to the cone stretching factor. Besides, Fourier expansion does not comply with an accurate tip-tilt filtering~\citep{Sasiela1994} and assume an infinite pupil that degrade the PSF model. Such considerations call for alternative approaches such as the point-wise method that is our formulation baseline described below.

\subsection{Point-wise calculation}

To maintain the accuracy of the anisoplanatism model whilst reducing the computational complexity, we focus on a point-wise approach. 
We follow the technique proposed by~\citep{Gilles2012}~: the phase is discretised over a grid of $N\times N$ pixels in the real domain, i.e. $\mathcal{M}_i(\rbb)$ functions are turned into Dirac distributions $\delta_i(\rbb) = \delta(\rbb - \rbb_i)$ that are centered at the $i^\text{th}$ pixel location. The anisoplanatic covariance takes the following form
\begin{equation} \label{E:covDelta_zonal}
\begin{aligned}
\cov{\Delta}(\rbb,\rhovec)
 = \sum_{i,j=1}^{N}\delta(\rbb-\rbb_i)\delta(\rbb + \rhovec - \rbb_i - \rhovec_j)\times\aver{\phi_\Delta(\rbb,t)\phi^t_\Delta(\rbb+\rhob,t)},
\end{aligned}
\end{equation}
where $\delta(\rbb-\rbb_i)\delta(\rbb + \rhovec - \rbb_i - \rhovec_j)$ is 1 only for the couple of pixels located at $\rvec_i$ and separated by $\rhovec_j$. Computation of $\cov{\Delta}(\rbb,\rhovec)$ is reduced to the determination of phase covariance at specific separations in the bi-dimensional plane. To include the dependency on both pupil location and separation, we compute the covariance of any two samples leading consequently $N^4$ values. All of these are concatenated into a $N^2\times N^2$ matrix which defines $\cov{\Delta}$. For a given separation, the latter is well described for the Von-K\'arm\'an spectrum of turbulence as
\begin{equation}\label{E:cov_fcn_zonal}
\begin{aligned}
	C_\phi(\rho) =  \left(\frac{L_{0}}{r_{0}}\right)^{5/3}\times
	\frac{\Gamma(11/6)}{2^{5/6}
		\pi^{8/3}}\times\left[\frac{24}{5}\Gamma\left(\frac{6}{5}\right)\right]\times
	\left(\frac{2\pi \rho}{L_{0}}\right)^{5/6}\times K_{5/6}\left(\frac{2 \pi \rho}{L_{0}}\right) 
\end{aligned}
\end{equation}
with $L_{0}$ and $r_{0}$ respectively the outer scale and  Fried's parameter, $\Gamma$ the 'gamma' function and finally $K_{5/6}$ a modified Bessel function of the third order.

Derivation of $\cov{\Delta}$ relies on the numerical implementation of Eq.~\ref{E:cov_fcn_zonal} that is fed with vector of separations to estimate the covariance terms in Eq.~\ref{E:covDelta_zonal}. The main challenge in the anisoplanatic covariance calculation lies in the proper definition of separations. 

\begin{figure}[h!]
	\centering
	\includegraphics[width=9cm]{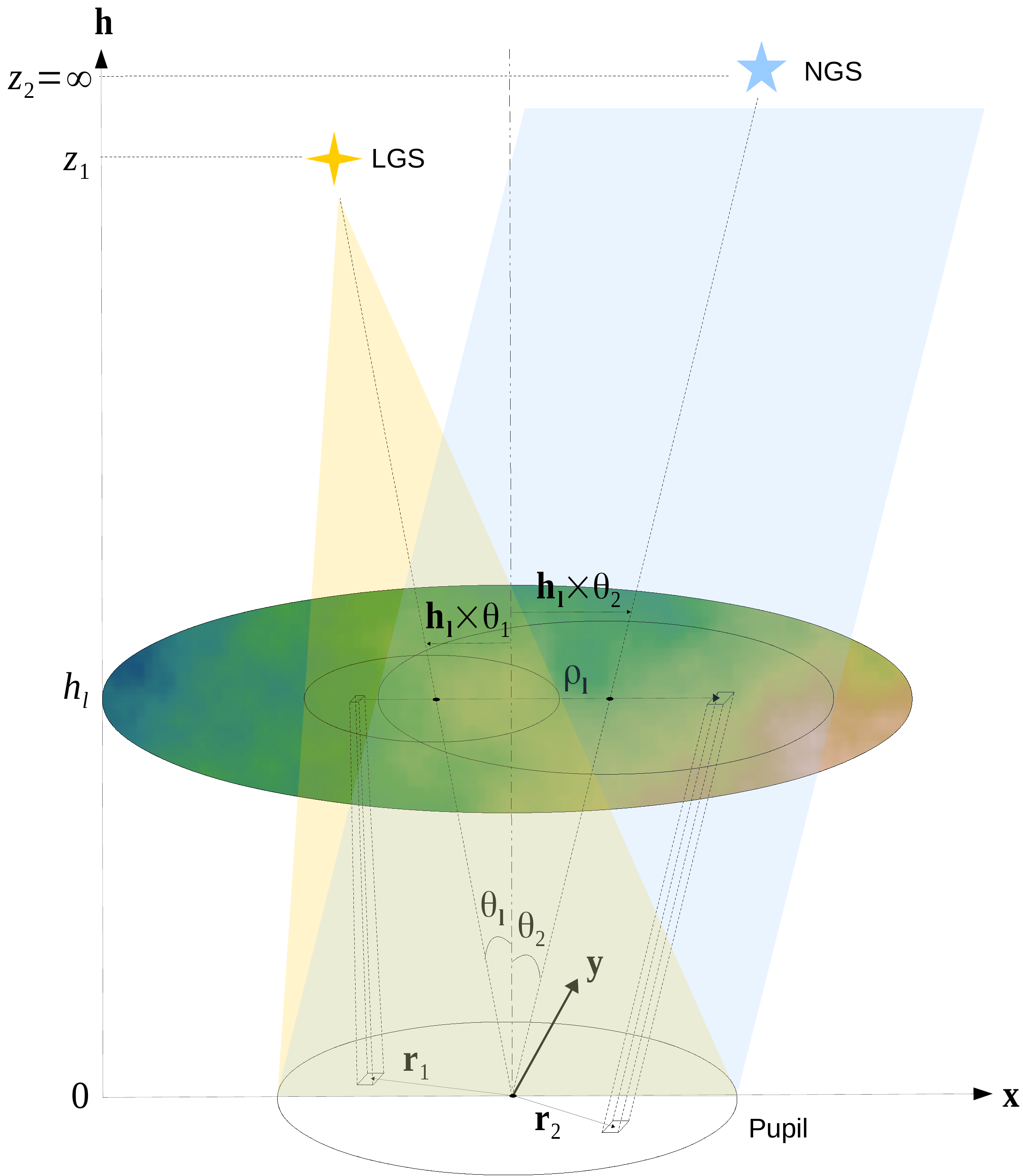}
	\caption{Sketch representing how the separation vector $\rhovec_l$ is defined between two resolution elements at a given turbulent layer $l$. Vectors $\rbun$ and $\rbdeux$ are the coordinates back-projected on the pupil.}
	\label{F:SACovMat}
\end{figure}

Consider the phase covariance along two different directions $\theta_1$ and $\theta_2$ at a single layer located at altitude $h_l$. Define $z_1$ and $z_2$ the source heights in direction 1 and 2 respectively. Finally, let $\rbb_1$ and $\rbb_2$ be the global pixel location vectors in the pupil projected from the atmosphere along straight paths. Fig.~\ref{F:SACovMat} provides a schematic view of this sketch. 
From Fig.~\ref{F:SACovMat}, we note the separation of two phase samples has two components. The first is related to the pupil plane coordinates that are stretched down with respect to cone-related squeeze factor $1-h_l/z_1$ and $1-h_l/z_2$. On top of that, there is a pupil shift in altitude created by the angular separation that leads to $h_l\times(\theta_1 - \theta_2)$. The separation vector $\rho_{l}(i,j)$, between aperture points raster index $i$ in direction $\thetavec_1$ and  index $j$ in direction $\thetavec_2$ at altitude $h_l$ is
\begin{equation}
\rho_{l}(i,j) = r_1(i)\times(1 - h_l/z_1) - r_2(j)\times(1 - h_l/z_2) + h_l\times(\theta_1 - \theta_2).
\label{E:rhodef}
\end{equation}
We assume the turbulence is discretised along $n_l$ statistically independent layers. From the last equation the element $(i,j)$ of $\cov{\Delta}$ takes the following expression
\begin{equation}
\begin{aligned}
	\cov{\Delta}(i,j) &= \sum_{l=1}^{n_l} f_l\times \left(\cov{\phi}((1 - h_l/z_1)r_1(i) - (1 - h_l/z_2)r_2(j)+ h_l\theta_1)\right.
		 +\cov{\phi}((1 - h_l/z_1)r_1(i) - (1 - h_l/z_2)r_2(j) + h_l\theta_2)\\
		 &- \cov{\phi}((1 - h_l/z_1)r_1(i) - (1 - h_l/z_2)r_2(j) + h_l\Delta\theta)
		 \left.- \cov{\phi}((1 - h_l/z_1)r_1(j) - (1 - h_l/z_2)r_2(i) + h_l\Delta\theta)\right),
\end{aligned}
\end{equation}
where $\Delta \theta = \theta_1 - \theta_2$, $f_l$ is the fractional power of the $l^\text{th}$ layer that is defined from the $\cnhl$ value as
\begin{equation}
	f_l = 0.06\lambda^2\rz^{-5/3}/\cnhl,
\end{equation}
where $\cnhl$ is the $\cnh$ profile at altitude $h_l$. This implementation of the ATF is freely made available with the simulator OOMAO~\citep{Conan2014OOMAO}. 
Other point-wise approaches have been developed~\citep{Tyler1994,Britton2006,Flicker2008}, but only the last reference is laser-compliant. For sanity check purpose, we have also coded the Flicker's method as an alternative point-wise calculation. Our formulation of anisoplanatism is strictly equivalent from the Flicker's one in the mathematical point of view; both of them relies on a non-stationarity calculation and include focal anisoplanatism. However we propose a direct derivation of the covariance function~(instead of the structure function as done by Flicker) that relies on optimal routines implementation within the OOMAO framework, that are 40~m class telescopes-compliant in terms of memory and cpu usage. 
	
\subsection{Generalization to laser-based systems}
\label{SS:lgsaniso}
In this section we provide the full expression of anisoplanatic covariance for laser-based systems which includes angular, focal and tilt terms. 

We define $\pas,\pal$ and $\pan$ as the atmospheric phase in respectively the science, LGS and NGS directions. The anisoplanatic phase $\phi_\Delta$ results from the subtraction of the high-order modes measured from the LGS and the tip-tilt measured on the NGS  from the phase in the science direction as
\begin{equation} \label{E:totaniso}
\phi_\Delta = \pas - \TTout.\pal - \TTonly.\pan
\end{equation}
where $\TTout(\rbb)$ is a spatial filter that removes the angle of arrival from LGS measurements. Conversely, $\TTonly(\rbb)$ is the filter that conserves only this angle of arrival. Modal spatial filters are of the kind
\begin{equation} 
    \TTout = \mathcal{I}_d - \TTonly
\end{equation}
with $\TTonly = \boldsymbol{T}\boldsymbol{T}^\dagger$ and $T$ the vector projecting the phase onto tip or tilt modes. Based on properties of $\TTonly$ and $\TTout$, $\phi_\Delta$ conforms to
\begin{equation}
\begin{aligned}
	\phi_\Delta &= \TTout\para{\pas - \pal} + \TTonly\para{\pas-\pan}\\
	& = \TTout\phi_{\Delta_\text{sl}} + \TTonly\phi_{\Delta_\text{sn}}	
\end{aligned}
\label{E:phiDelta_lgs}
\end{equation}
where a first term $\TTout\phi_{\Delta_\text{sl}}$ relates to the anisoplanatism, both angular and focal whereas the second term $\TTonly\phi_{\Delta_\text{sn}}$ represents tilt anisoplanatism. From Eq.~\ref{E:phiDelta_lgs}, the anisoplanatic covariance $\cov{\Delta}$ in Eq.~\ref{E:cphi0} becomes
	\begin{equation} 
	\begin{aligned}
	C_{\Delta}(\rbb,\rhovec)  & = \TTout\aver{\phi_{\Delta_\text{sl}}(\rbb)\phi^t_{\Delta_\text{sl}}(\rbb+\rhovec)}\TTout^t\\
	& + \TTonly\aver{\phi_{\Delta_\text{sn}}(\rbb)\phi^t_{\Delta_\text{sn}}(\rbb+\rhovec)}\TTonly^t\\
	& + \TTonly\aver{\phi_{\Delta_\text{sn}}(\rbb)\phi^t_{\Delta_\text{sl}}(\rbb+\rhovec)}\TTout^t\\
	& + \TTout\aver{\phi_{\Delta_\text{sl}}(\rbb)\phi^t_{\Delta_\text{sn}}(\rbb+\rhovec)}\TTonly,	 
	\end{aligned}
	\label{E:cov2}
	\end{equation}
which turns into
	\begin{equation} 
	\begin{aligned}
	C_{\Delta}(\rbb,\rhovec) & =  \TTout\cov{\Delta_\text{sl}}(\rbb,\rhovec)\TTout^t\\
	&+ \TTonly\cov{\Delta_\text{sn}}(\rbb,\rhovec)\TTonly^t \\
	& + \cov{\text{cross}}(\rbb,\rhovec) + \cov{\text{cross}}^t(\rbb,\rhovec),
	\end{aligned}
	\label{E:cov3}
	\end{equation}
where covariance terms $\cov{\Delta_\text{sl}}$ and  $\cov{\Delta_\text{sn}}$ are respectively focal-angular and tip-tilt anisoplanatism that are derived independently using the formalism introduced in the previous section. Eq.~\ref{E:cov3} introduces cross-correlation on high-order and tip-tilt modes. From Eqs.~\ref{E:cov2} and~\ref{E:phiDelta_lgs}, this cross term writes
\begin{equation}
\begin{aligned}
	\cov{\text{cross}}(\rbb,\rhovec) &= \TTonly\aver{(\pas(\rbb) - \pan(\rbb))(\pas(\rbb+\rhovec)-\pal(\rbb+\rhovec))^t}\TTout^t\\
	& = \TTonly\left(\aver{\pas(\rbb)\pas(\rbb+\rhovec)^t} + \aver{\pan(\rbb)\pal^t(\rbb+\rhovec)}\right.\\
	& - \left.\aver{\pas(\rbb)\pal^t(\rbb+\rhovec)} + \aver{\pan(\rbb)\pas^t(\rbb+\rhovec)}\right).\TTout^t,
\end{aligned}
\end{equation}
and includes the tip-tilt/high order modes correlation. These cross-terms are generally neglected in the literature on account of its small amplitude when comparing with other terms; we propose to confirm this assumption using full physical-optics simulations in Sect.~\ref{SS:lgscase}. We now introduce the terminology \emph{total anisoplanatism} when including all these cross-terms as in Eq.~\ref{E:cov3}, compared to the {split anisoplanatism} that considers $\cov{\text{cross}} = 0$ in Eq.~\ref{E:cov3}.

\subsection{Spatial filtering}

The number of phase samples $N$ must be chosen wisely; large values will make the calculation time-consuming, while  small values will make us loose in accuracy. In terms of spatial frequencies, $N$ phase samples allows to derive the PSF within a $N/2\times \lambda/D$-wide 2D region. Besides, the AO correction band breaks at $n_\text{act}/2\times \lambda/D$, with $n_\text{act}$ the linear number of DM actuators, where the anisoplanatism occurs actually. As a consequence, it is enough to define the $\cov{\Delta}$ as a $n_\text{act}\times n_\text{act}$ matrix to represent accurately the anisoplanatism in the AO correction band.  However numerical simulations require to sample the phase with a resolution given by at least 2-3 points per $\rz$, which potentially makes $N > n_\text{act}$. We solve this issue by multiplying $\cov{\Delta}$ with a filter matrix $\Hdm$ that is designed to filter out uncontrolled DM modes; it sets to zero all spatial frequencies greater than $n_\text{act}/2D$
\begin{equation} \label{E:atf_filt}
\begin{aligned}
\atf{\Delta}(\rhovec/\lambda) = \dfrac{1}{\otf{\text{DL}}(\rhovec/\lambda)}\times\iint_{\mathcal{P}}  \mathcal{P}(\rbb)\mathcal{P}(\rbb+\rhovec)\times
\exp\para{\Hdm\para{\cov{\Delta}(\rbb,\rhovec) - \cov{\Delta}(\rbb,0)}\Hdm^t} \boldsymbol{dr}
\end{aligned}
\end{equation}
where $\Hdm$ is a zonal DM filter of size $N\times N$, calculated from the DM influence functions $\boldsymbol{h}$ of size $N^2\times n_\text{act}$ as follows
\begin{equation}
	\Hdm = \mathcal{I}_d - \boldsymbol{h}\boldsymbol{h}^\dagger
\end{equation}
where $\mathcal{I}_d$ is the $N\times N$ identity matrix. This filter removes spatial frequencies above the DM cut-off frequency that are beyond the correction space spanned by the latter. The removed high spatial frequencies contribute to the fitting error; it manifests itself as the PSF halo and is a function of the atmospheric seeing only.

\section{Analytic models versus physical-optics simulations}
\label{S:comparison}

In this section we compare existing anisoplanatism models in the literature to simulations. In a first step, we aim at checking the point-wise derivation of anisoplanatism using our general formalism, refereed as OOMAO in next results, complies with existing models, as Zernike~\citep{Fusco2000}, Fourier~\citep{VanDam2006} and~\citep{Flicker2008}. 

\subsection{Metrics}
\label{S:metrics}

Results presented in the next sections evaluate PSF model deviations to simulation.  First insights on PSF are given by morphological scalar values, as the Strehl-ratio and the FWHM. Both these parameters mostly refer to AO performance; we might prefer having a metric that encompasses the entire structure of the PSF, including especially the PSF halo outside the AO-correction band, as the Fraction of Variance Unexplained~(FVU). If $X$ is 2-D image and $\widehat{X}$ the estimation of this, the FVU is defined as\citep{King1983}
\begin{equation} 
\text{FVU}_X = \dfrac{\sum_{i,j} \para{X(i,j) - \widehat{X}(i,j)}^2}{\sum_{i,j} \para{X(i,j) - \sum_{i,j}X(i,j)}^2},
\label{E:FVU}
\end{equation} 
where $(i,j)$ are respectively the $\xth{i}$ and $\xth{j}$ pixel of the image. The great advantage of such a metric is to yield an overall error on all angular frequencies; all the PSF patterns, as the PSF core and wings, are included into this calculation.

Besides PSF-related metrics, characterising off-axis PSFs is motivated by science exploitation, calling for science-based metrics, such as photometry and astrometry. Contrary to PSF-metrics, science ones must refer to a specific observed object and image processing tools~(deconvolution or model-fitting for instance). We focus on a particular science case of imaging a binary system to measure relative fluxes and astrometry for deriving the binary’s orbit. In this case, the field typically lacks independent PSF stars and the ability to characterize the PSF is of tremendous value. We define a binary model from a reference off-axis PSF and a set of parameters as
\begin{equation}
\begin{aligned}
\mathcal{B}\para{\text{PSF},\Delta F,\Delta \alpha_x,\Delta \alpha_y} =\Delta F\times\para{\text{PSF}\para{\alpha_x,\alpha_y}\ + \text{PSF}\para{\alpha_x + \Delta \alpha_x ,\alpha_y+\Delta \alpha_y}}
\end{aligned}
\end{equation}
where $(\alpha_x,\alpha_y)$ represents the angular separations in the focal plane, $\Delta F$ the relative stars flux and ($\Delta \alpha_y$,$\Delta \alpha_y$) the differential angular offsets. We did not include any source of noise and AO residual in the focal plane to really extract the real impact of anisoplanatism characterization onto our metrics.

We define a reference binary model from the simulated off-axis PSF, sampled at $\lambda/4D$, with $\Delta F^0 = 1$ and a star separation set to $\Delta\alpha_y^0 = \lambda/D$. Accuracy on photometry and astrometry is evaluated by minimizing the following criterion
\begin{equation} \label{E:cost}
\begin{aligned}
\varepsilon^2(\Delta F,\Delta\alpha_x,\Delta\alpha_y) =\norme{\mathcal{B}\para{\text{PSF}_\varepsilon,\Delta F^0,\Delta\alpha_x^0,\Delta\alpha_y^0} - \mathcal{B}\para{\text{P}\widehat{\text{S}}\text{F}_\varepsilon,\Delta F,\Delta \alpha_x,\Delta \alpha_y}}^2_2
\end{aligned}
\end{equation}
where $\norme{\boldsymbol{x}}_2^2$ is the $\mathcal{L}_2$ norm of the vector $\boldsymbol{x}$. We retrieve photometry and astrometry by fitting a synthetic PSF-based binary on the reference model. Because we know the exact binary parameters, we can estimate how we deviate from those regarding the PSF model. Particularly the photometry error is given by
\begin{equation}
\Delta \text{mag} = -2.5\times\log_\text{10}\para{\dfrac{\widehat{\Delta}F}{\Delta F^0}}
\end{equation}
while the astrometry error results from
\begin{equation}
\Delta \alpha = \sqrt{\para{\widehat{\Delta}\alpha_x - \Delta\alpha_x^0}^2 + \para{\widehat{\Delta}\alpha_y -\Delta\alpha_x^0}^2}.
\end{equation}

Photometry and astrometry, as defined in last equations, may be derived differently regarding the science case and the image-processing method. However, gathering science and PSF-related metrics in our analysis will enhance the overall evaluation of anisoplanatism characterization; we will point out the consequence of PSF errors on science images exploitation, that is the real information that matters in the end. Finally, we estimate astrometry accuracy with an infinite signal-to-noise ratio; we consider only PSF morphology impact into this derivation.

\subsection{NGS case}

We have simulated H-band atmospheric phase screens using the simulator OOMAO~\citep{Conan2014OOMAO} to compare all anisoplanatism models described in the previous section to simulations. These latter include only anisoplanatism patterns into the PSF; we do not account for AO-residual in the guide star direction, static patterns ans science camera parasites such as noise. Our goal is strictly dedicated to anisoplanatism determination and impact on the PSF.

\begin{figure}
	\centering
	\includegraphics[height=9cm]{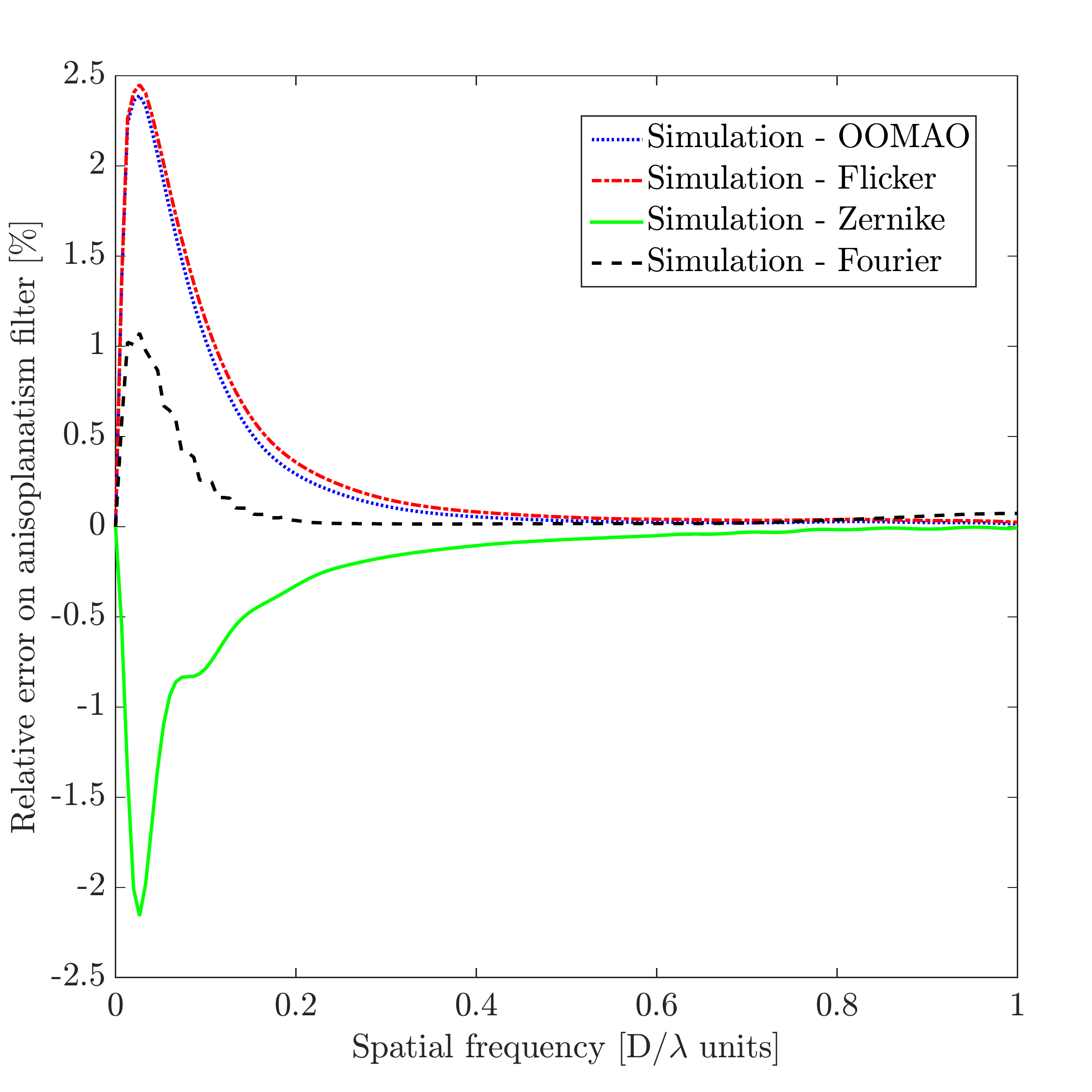} 
	\caption{\small Relative errors on ATF azimuthal average deduced from simulation and analytical calculations difference, for a NGS 40"-off. Negative values on residuals corresponds to overestimation of analytical calculation.}
	\label{F:atf0}
\end{figure}

We have considered the median $\cnh$ profile~\citep{Sarazin2013} at Paranal degraded to 7-layers with $\rz = 16$ cm. See next section for explanation about the choice of a 7-layers based profile. Fig.~\ref{F:atf0} provides a qualitative comparison and illustrates how each approach achieves an accurate model of the ATF within a percent of the full physical-optics simulation. We notice the Zernike approach becomes less accurate when putting the NGS farther away in the field. We may believe that the stationarity assumption causes this effect, as invoked by~\citep{Fusco2000}, but in such a case, we would observe a similar degradation compared to the Fourier approach that relies on the same assumption as well. The main issue here is that the DM spatial filtering is translated into a modal truncation at the radial order $n$ given by $0.3\times(n+1)/D$~\citep{Conan1994}. Zernike modes do not match exactly the DM frequency behaviour; such an approximation in the DM cut-off frequency may introduce slight errors on the ATF, especially when the anisoplanatism level is stronger. The Fourier method is the most accurate for the NGS case in the FVU sense despite the approximation made of infinite pupil translating into potential edge effects. 

Tab.~\ref{T:model} provides quantitative assessment of ATF compared to simulations and highlights each method leads to very similar results. What we see is consistent with Fig.~\ref{F:atf0}: all models match very well to end-to-end simulations, within 0.2\% of FVU on OTF. We also verified that FVU on PSF are systematically within one or two order of magnitude lower than FVU on ATF. It is explained by the PSF derivation that results from the Fourier-transform of the ATF, that is preliminary multiplied by the diffraction-limit OTF, given by Eq.~\ref{E:otfDL}, that filtered out high-angular frequencies.  Strehl ratio is estimated within few percent, which is definitely below errors bars obtained on images acquired on-sky, while FWHM is perfectly well estimated whatever the approach. Science metrics reach a level of milli-mag~(0.1\% on photometry) and micro-arcsec~(1\% of pixel-size), that is definitely at several order of magnitude lower than usual estimations~\citep{Turri2017,Ascenso2015}. 

\begin{table*}
	\centering
	\captionof{table}{\small Relative errors obtained on considered PSF metrics by comparing in simulated anisoplanatic PSF to modeled ones. Models refer respectively to OOMAO, Flicker, Zernike and Fourier. Photometric errors are given in H-band milli-mag and astrometry in $\mu$as. Zero values corresponds to machine precision. Strehl-ratio and FWHM given on table header are values extracted out the simulation.}
	\small
	\begin{tabular}{|c|c|c|c|c||c|c|c|c||c|c|c|c|}				
		\hline
		NGS location & \multicolumn{4}{c|}{{10"}}  & \multicolumn{4}{c|}{{20"} } & \multicolumn{4}{c|}{{40"}} \\
		\hline
		Strehl-ratio [\%] & \multicolumn{4}{c|}{54}  & \multicolumn{4}{c|}{19} & \multicolumn{4}{c|}{5}\\
		\hline
		FWHM [mas] &\multicolumn{4}{c|}{80}  & \multicolumn{4}{c|}{90} & \multicolumn{4}{c|}{260} \\
		\hline
			& \multicolumn{12}{c|}{Relative residual errors [\%]} \\
		\hline		
		Model 	& O & F & Ze & Fo & O & F & Ze& Fo& O& F& Ze&Fo \\		
			\hline
		SR 		&3.8&4.0&1.1 &2.3 & 3.4& 3.9&1.7 &1.6 & 2.7& 3.3& 3.5&0.6 \\			
		\hline
		FWHM &  0&  0& 0.1 &0.1  & 0.04& 0.04 & 0.2& 0.6& 0.4& 0.06& 1.0&1.5 \\				
		\hline		
		FVU$_\text{OTF}$ & 0.21 &0.23  & 0.02 & 0.08 & 0.15&0.18  & 0.04& 0.05& 0.08& 0.1& 0.08& 0.03\\		
		\hline
		& \multicolumn{12}{c|}{Science estimates} \\
		\hline
		$\Delta$mag 	& 25 &29  &2  &13  & 22& 28 &4 & 6& 15&20 &20 &5 \\		
		\hline
		$\Delta \alpha $ 	& 2 & 9 & 37 &28  &13 &49 &56 &50 & 50& 10& 14&70 \\		
		\hline
	\end{tabular}	
	\label{T:model}
\end{table*}
As a conclusion, our point-wise method matches accurately existing angular anisoplanatism models and simulations in the literature within 1\%-level differences.

\subsection{LGS case}
\label{SS:lgscase}

We now focus on anisoplanatism on LGS-based systems; we have simulated the total anisoplanatism from Eq.~\ref{E:totaniso}, with one LGS and one NGS distributed along a L-shaped asterism while keeping the science on-axis. We have estimated the anisoplanatic covariance given in Eq.~\ref{E:cov2} and the resulting ATF using Eq.~\ref{E:atf}. Furthermore, we have also derived independently focal and tilt anisoplanatic covariance matrices $\cov{\Delta_{sl}}$ and $\cov{\Delta_{sn}}$ in Eq.~\ref{E:cov3}, using both simulations and analytic calculations. The goal of the analysis is two-fold: firstly demonstrate the generalized analytic model matches simulations including all anisoplanatism terms, secondly confirming high order/tip-tilt cross terms that appear in Eq.~\ref{E:cov2} are not determinant in the anisoplanatism characterization. 

\begin{figure}
		\centering
		\includegraphics[height=9cm]{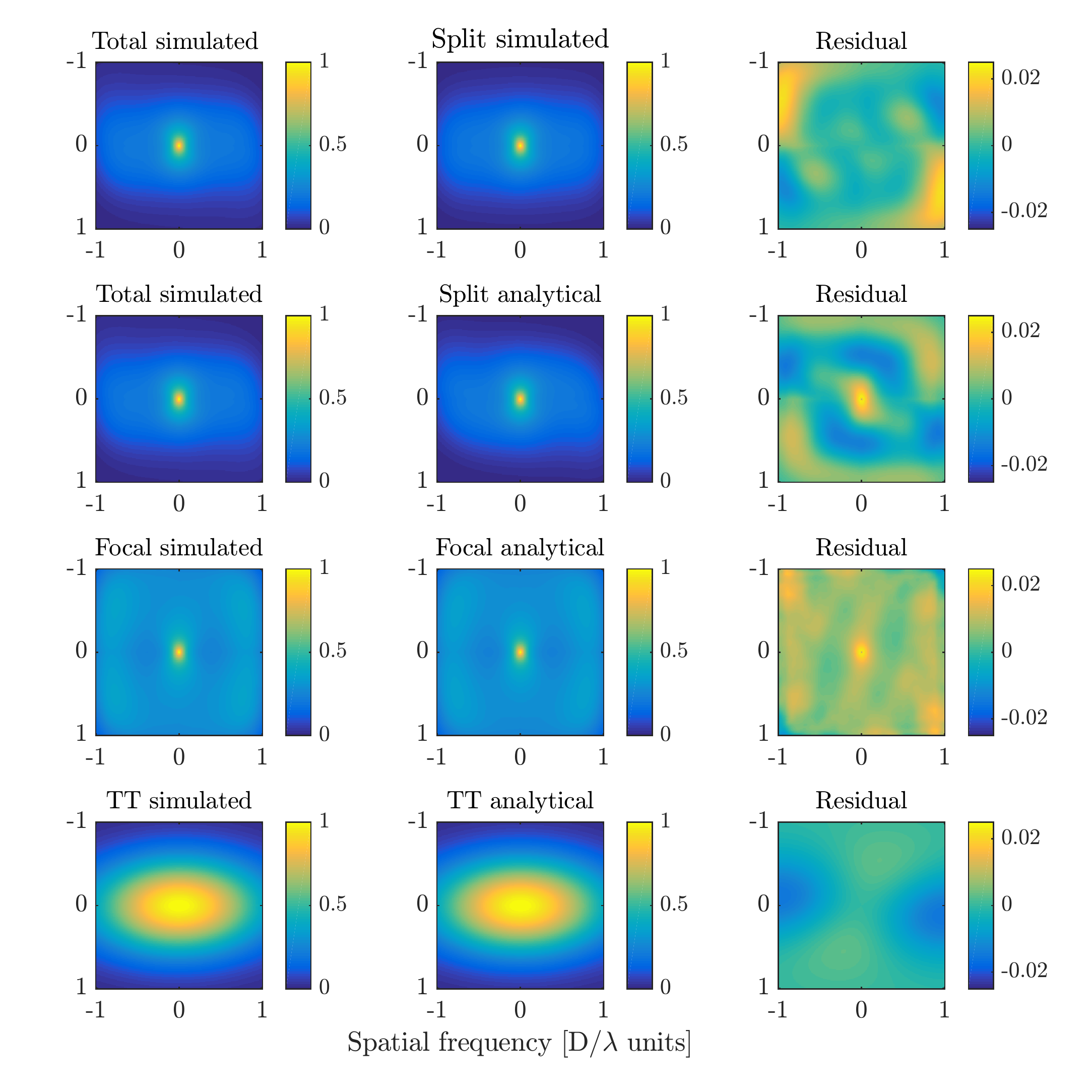} 
		\captionof{figure}{\small Comparison of ATF maps derived either from simulation and analytical calculations. Figure distinguishes total and split anisoplanatism that respectively do and do not include tip-tilt/high order modes cross-correlation as discussed in Sect.~\ref{SS:lgsaniso}. ATF maps are derived for a LGS and NGS 40"-off respectively to the north and east. }
		\label{F:atfs}
\end{figure}

\begin{figure}
	\centering	
	\includegraphics[height=9cm]{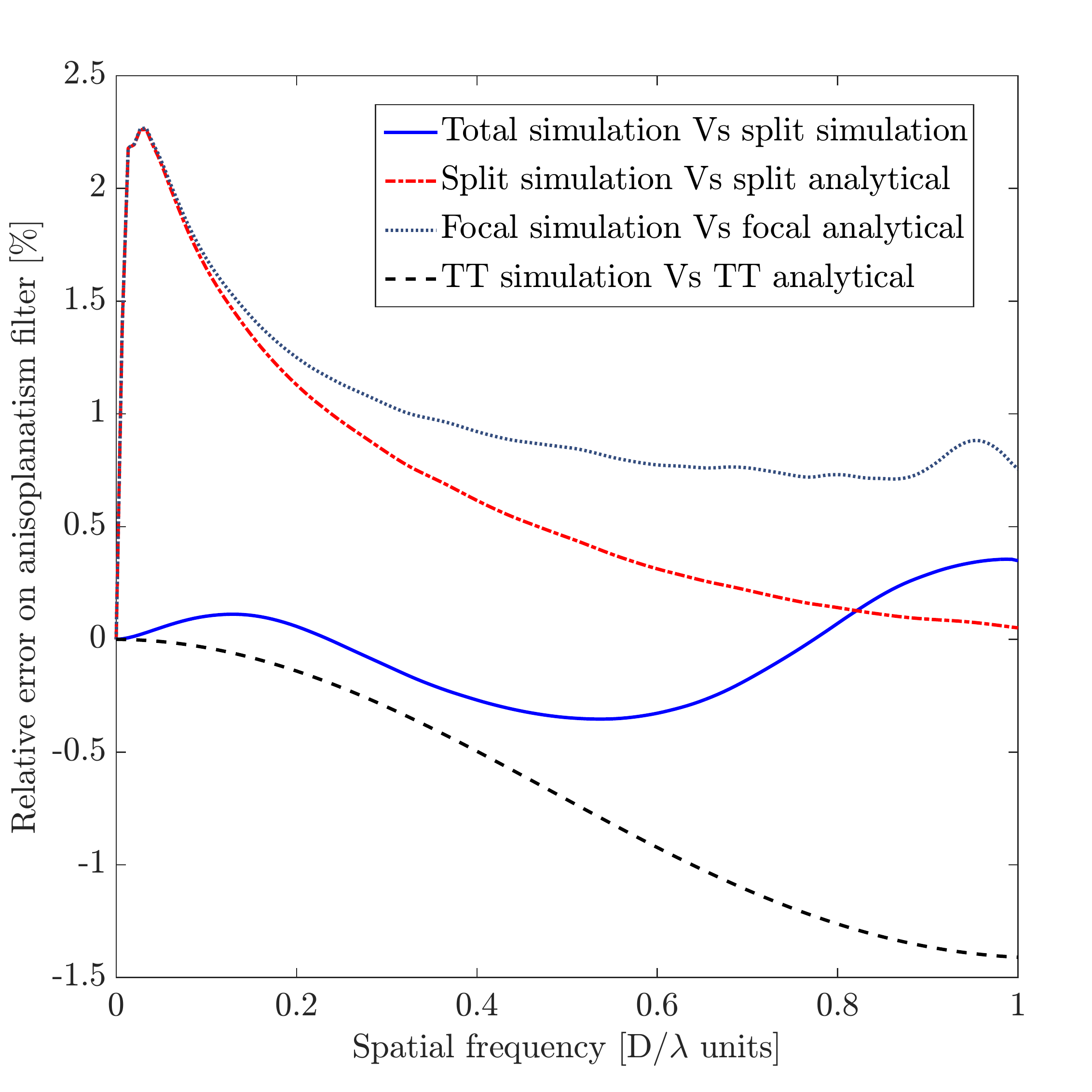}
	\caption{\small Relative errors on ATF azimuthal average deduced from simulation and analytical calculations difference for an LGS at 20" and NGS at 40" separation respectively to the north and east.}
	\label{F:splitVtotal}
\end{figure}

Fig~\ref{F:atfs} provides a comparison of ATFs maps for a 20"-off LGS and 40"-off NGS from on-axis in perpendicular directions. It highlights clearly that analytic calculations fit very well simulations results to within two percent points of accuracy. On top of that, residual errors are mostly located at the map border that corresponds to pupil edges. As previously, FVU on PSF reach two order of magnitude lower level compared to FVU on OTF. When extrapolating the on-axis OTF, we will multiply these angular frequencies above $D/\lambda$ to zero. Fig~\ref{F:splitVtotal} illustrate cuts of ATF and residuals along the elongated direction. We observe that analytic calculations reproduce very well simulations within 2\% in maximal range and 0.1\% in FVU when looking at Tab.~\ref{T:simuvzonal}. 

\begin{table}
	\centering
	\captionof{table}{\small Relative errors obtained on considered PSF metrics by comparing in simulated PSFs to analytical ones considering split anisoplanatism. Sources were distributed along a L-shape constellation with the science target on-axis. Photometric errors are given in H-band milli-mag and astrometry in $\mu$as. Zero values corresponds to machine precision. Strehl-ratio and FWHM given on table header are values extracted out the simulation.}
	\small
	\begin{tabular}{|c||c|c|c||c|c|c|}
		\hline
		LGS location & \multicolumn{3}{c||}{0"} & \multicolumn{3}{c|}{20"} \\
		\hline
		NGS location & 0"  & 20"  & 40"  & 0" & 20" & 40" \\
		\hline
		Strehl-ratio [\%] & 64  &56  & 41 &  22  & 19 & 15\\
		\hline
		FWHM [mas] & 79 &84  & 97 & 86 &92  & 108\\
		\hline
		& \multicolumn{6}{c|}{Relative residual errors [\%]} \\
		\hline		
		SR& 2.2 & 2.0 & 2.0 & 2.5 & 2.03 & 2.2  \\			
		\hline
		FWHM & 0 & 0.31&0.35& 0 &  0.23  & 1.3 \\			
		\hline		
		FVU$_\text{OTF}$ & 0.07 & 0.06 & 0.06 & 0.10 &0.08 & 0.08 \\		
		\hline
		& \multicolumn{6}{c|}{Science estimates} \\
		\hline
		$\Delta$mag &  6.3  & 3.1 & 4.8 &  8.6 &  1.8  &  6.0  \\
		\hline
		$\Delta \alpha $ & 7.7 & 41.4 & 28.2 & 67.6 & 60.5  & 77.6  \\	
		\hline
	\end{tabular}	
	\label{T:simuvzonal}
\end{table}

We went through systematic comparisons over different asterism geometries to investigate whether cross terms are always negligible. Tab.~\ref{T:totvsplit} highlights that cross-terms independence assumption is affecting the PSF within extremely low differences, lower than model deviation from simulations. As a sub-product of our analysis, we confirm the LGS-based anisoplanatism is accurately represented by two independent terms, as angular+focal terms and tilt anisoplanatism.  
\begin{table}
	\centering
	\captionof{table}{\small Relative errors obtained on considered PSF metrics by comparing in simulation PSFs produced by the total and split anisoplanatism. Sources were distributed along a L-shape constellation with the science target on-axis. Photometric errors are given in H-band milli-mag and astrometry in $\mu$as. Zero values corresponds to machine precision.}
	\small
	\begin{tabular}{|c||c|c|c||c|c|c|}
		\hline
		LGS location & \multicolumn{3}{c||}{0"} & \multicolumn{3}{c|}{20"} \\
		\hline
		NGS location & 0" & 20" & 40" & 0"& 20"& 40" \\
		\hline
		& \multicolumn{6}{c|}{Relative residual errors [\%]} \\
		\hline		
		SR& 0 & 3e-2 & 3e-2 & 0 & 2e-2 & 1e-2  \\			
		\hline
		FWHM & 0 & 0.35& 0.39& 0.22 & 0.39 & 1.39 \\			
		\hline		
		FVU$_\text{OTF}$ & 0 & 1e-4& 5e-4& 0 &2e-3 & 4e-3 \\		
		\hline
		& \multicolumn{6}{c|}{Science estimates} \\
		\hline
		$\Delta$mag &0&0.30 &0.21 &2e-3 & 1.05 & 1.5  \\		
		\hline
		$\Delta \alpha $ & 0& 2.0 & 16.7& 0.05&8.1 &36.0  \\	
		\hline
	\end{tabular}	
	\label{T:totvsplit}
\end{table}

\section{Off-axis PSF sensitivity to $\boldsymbol{\cnh}$ profile accuracy}
\label{S:sensitivity}
In this section we quantify the impact of $\cnh$ profile knowledge on PSF morphological and science metrics. 

We split our study into two complementary analyses. On a first step, we introduce a bias on the $\cnh$ estimation by binning down the number of layers. On a second step, we keep the same number of bins, but we introduce random variations on both heights and weights.

\subsection{Impact of number of bins}

We have initially considered a 35-layers $\cnh$ profile~\citep{Sarazin2013} as our reference to compute the PSF at different separations ($\theta_0/2$, $\theta_0$, $1.5\times\theta_0$ and $2\times\theta_0$) with $\theta_0$ = 24.5" in H-band. 

We have opted for the mean-weighted compression~\citep{Robert2010} method for reducing the problem from 15 down to 2 layers. At each iteration, we retrieve the anisoplanatic PSF to compare to the reference full profile PSF. Other binning options could be considered instead~\citep{Saxenhuber2017}. Keeping the angular coherence angle constant across profiles seems to us a sensible choice as we look particularly into anisoplanatic effects.  

\begin{figure} 
	\centering
	\includegraphics[height=9cm]{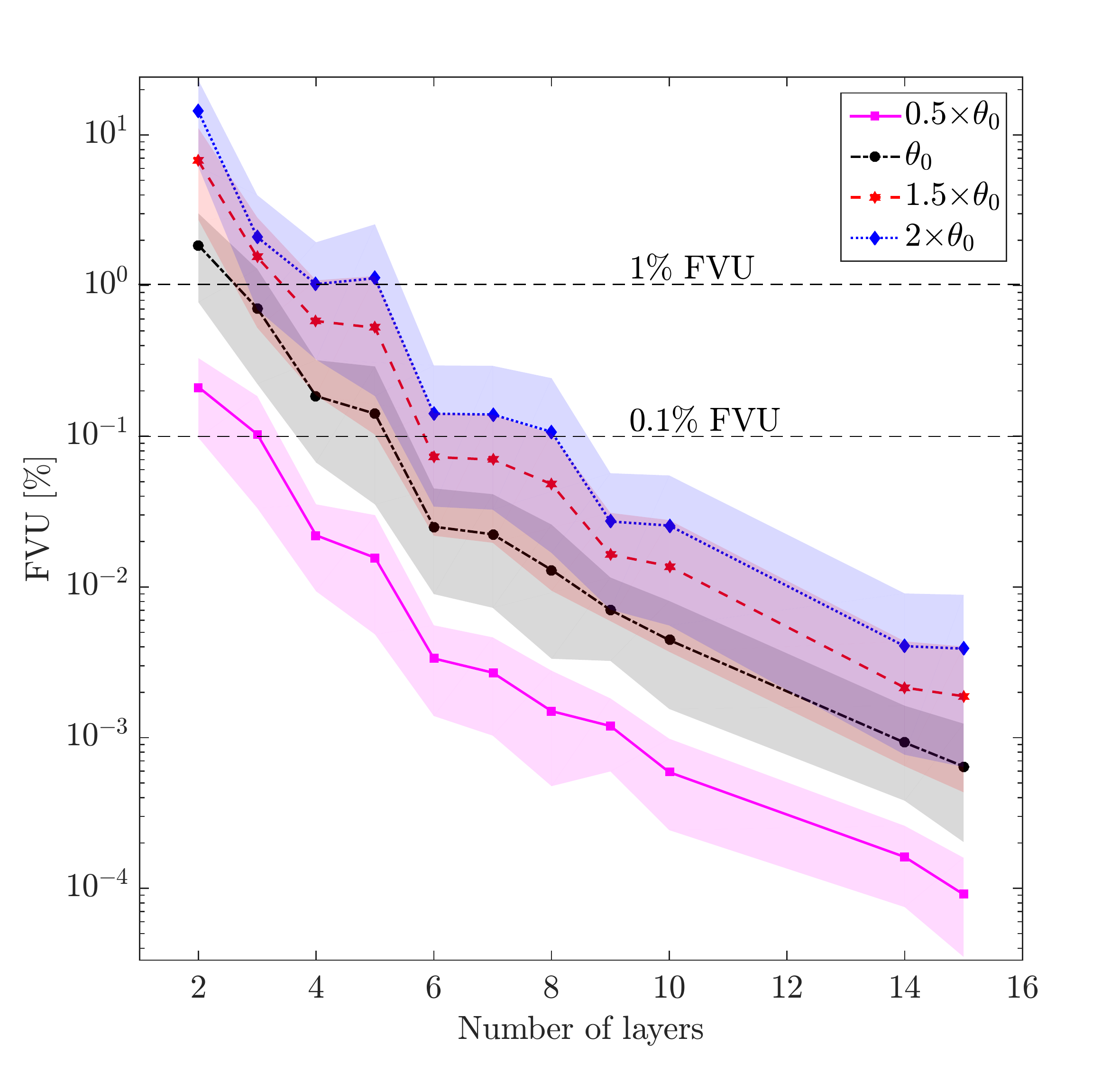}
	\captionof{figure}{\small  Fraction of variance unexplained as function of the number of reconstructed layer.}
	\label{F:fvuVnlayer}
\end{figure}

\begin{figure} 
	\centering
	\includegraphics[height=8cm]{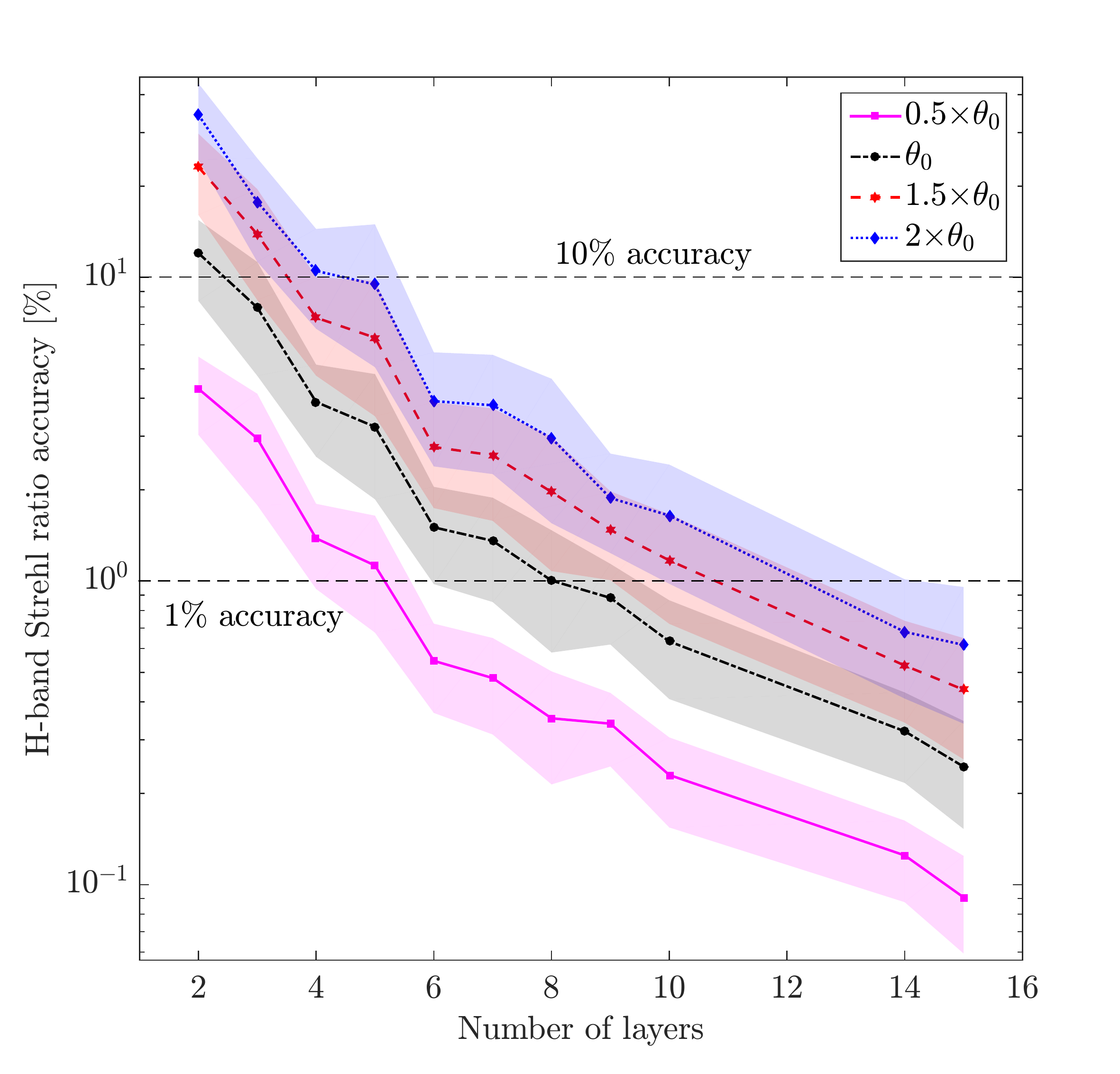}
	\includegraphics[height=8cm]{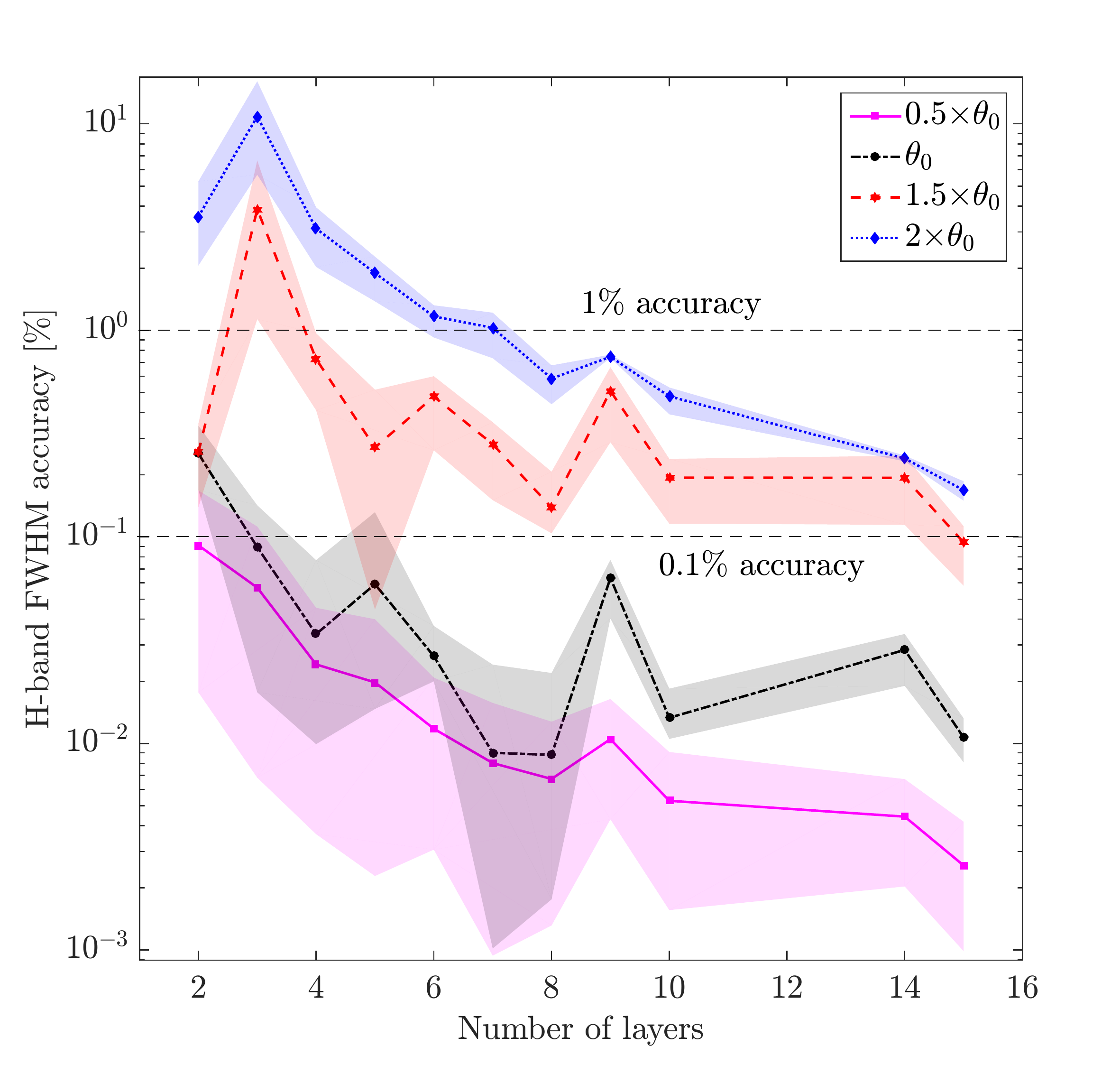}
	\captionof{figure}{\small \textbf{Left~:} H-band long-exposure Strehl ratio \textbf{Right~:} FWHM accuracy versus the number of reconstructed layer.}
	\label{F:srfwhmVnLayer}
\end{figure}

\begin{figure} 
	\centering
	\includegraphics[height=8cm]{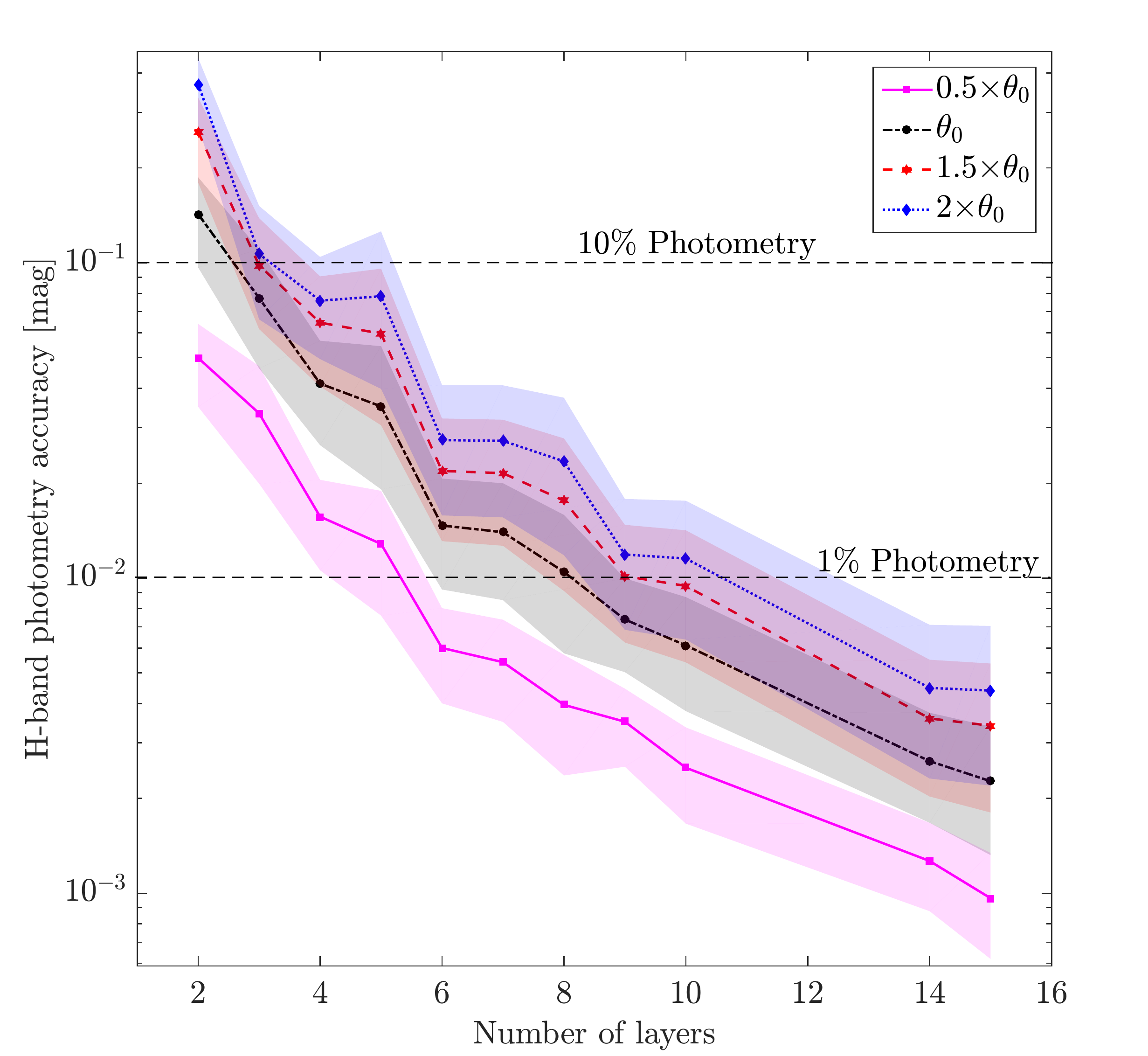}
	\includegraphics[height=8.25cm]{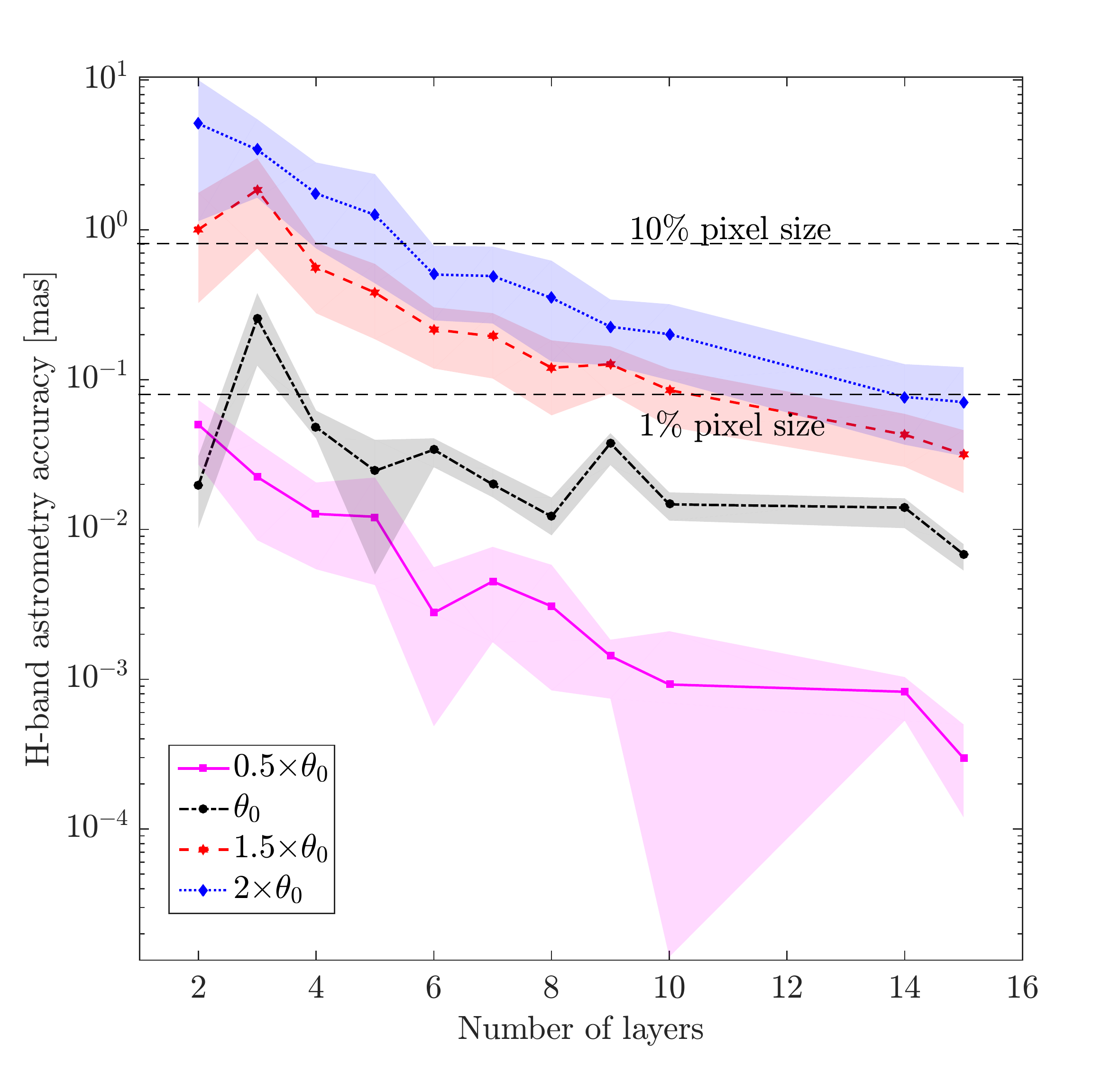}
	\captionof{figure}{\small \textbf{Left~:} H-band photometry \textbf{Right~:} Astrometry	accuracy versus the number of reconstructed layer.}
	\label{F:photoastroVnLayer}
\end{figure}

We report on Figs.~\ref{F:fvuVnlayer},~\ref{F:srfwhmVnLayer} and~\ref{F:photoastroVnLayer} the FVU and accuracy on Strehl-ratio, FWHM, photometry and astrometry as function of the number of modelled layers for a median profile at Paranal and different off-axis position in the field~($0.5\times\theta_0$,$\theta_0$,$1.5\times\theta_0$ and $2\times\theta_0$). Curves envelopes are deduced from quartiles profiles~\citep{Sarazin2013}.
An immediate observation is that 15 layers instead of 35 can be used with little impact on the PSFs. If 1\% errors are allowed, at least 7 layers are required. For such a profile, photometry errors are at the level of 3\%-level while  astrometry is given at 5\% of pixel size level in the worst case. External profilers commonly deliver profiles with such a number of layers; these results suggest that model-dependent errors based on that input are of the percent level. 

 This is an important take home message for anisoplanatism characterization and PSF-R on 10~m class telescopes: for a field roughly given by $\theta_0$, we only need a representation of the $\cnh$ over 7-layers to model anisoplanatism signature onto PSF at 1\%-level. 

\subsection{Impact of heights and weights precision}

We consider the 7-layer equivalent profile at Paranal~\citep{Sarazin2013} as the new reference for this section; our purpose is to evaluate how much the precision on both weights and heights of turbulent layers impact the PSF. We denote $h_l^0$ and $w_l^0$ respectively to present the $l^\text{th}$ layer heights and weights values of reference.

We apply a random variation on each layer, either on its weight $w_l$ or height $h_l$. We define for each of those layers a zero-mean Gaussian statistical variable - $\eta_h$ and $\eta_w$ -, by setting its standard-deviation to unity. We then define $\sigma_h$ and $\sigma_w$ as $p$-size vectors such as
\begin{equation}
\begin{aligned}
&h_l(p,k) = h_l^0 +  \sigma_h(p)\times\eta_h(l,k)\\
& w_l(p,k) = w_l^0\times\para{1 + \sigma_w\times\eta_w(l,k)}
\end{aligned}
\end{equation}
where $k$ refers to the random selection and $p$ to the value of deviation introduced. For height sensitivity, we allow up to 1km of variability, while weights are de-tuned by up to 30\%. For each iteration $k$, we define a new 7-layer atmosphere and get the PSF at any separation starting from the PSF on-axis as described earlier. We finally compute metrics as a function of $\sigma_h$ and $\sigma_w$, the separation $\theta$ and the random selection $k$. Metrics are averaged out over 1000 realizations providing error bars as well. The choice of 1000 iterations is a compromise between the computation time and relevance of results.

\begin{figure}
	\centering
	\includegraphics[height=9cm]{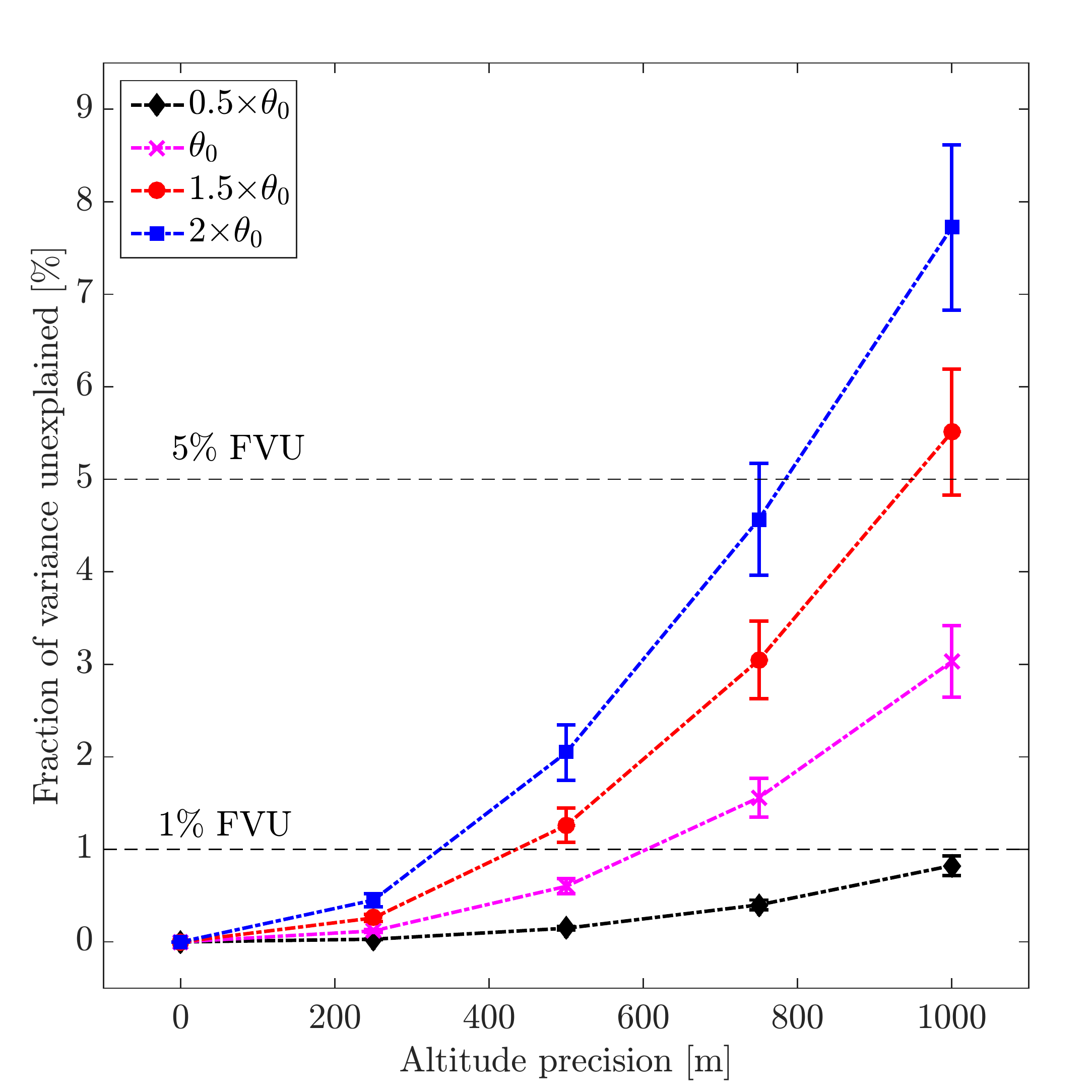}
	\includegraphics[height=9cm]{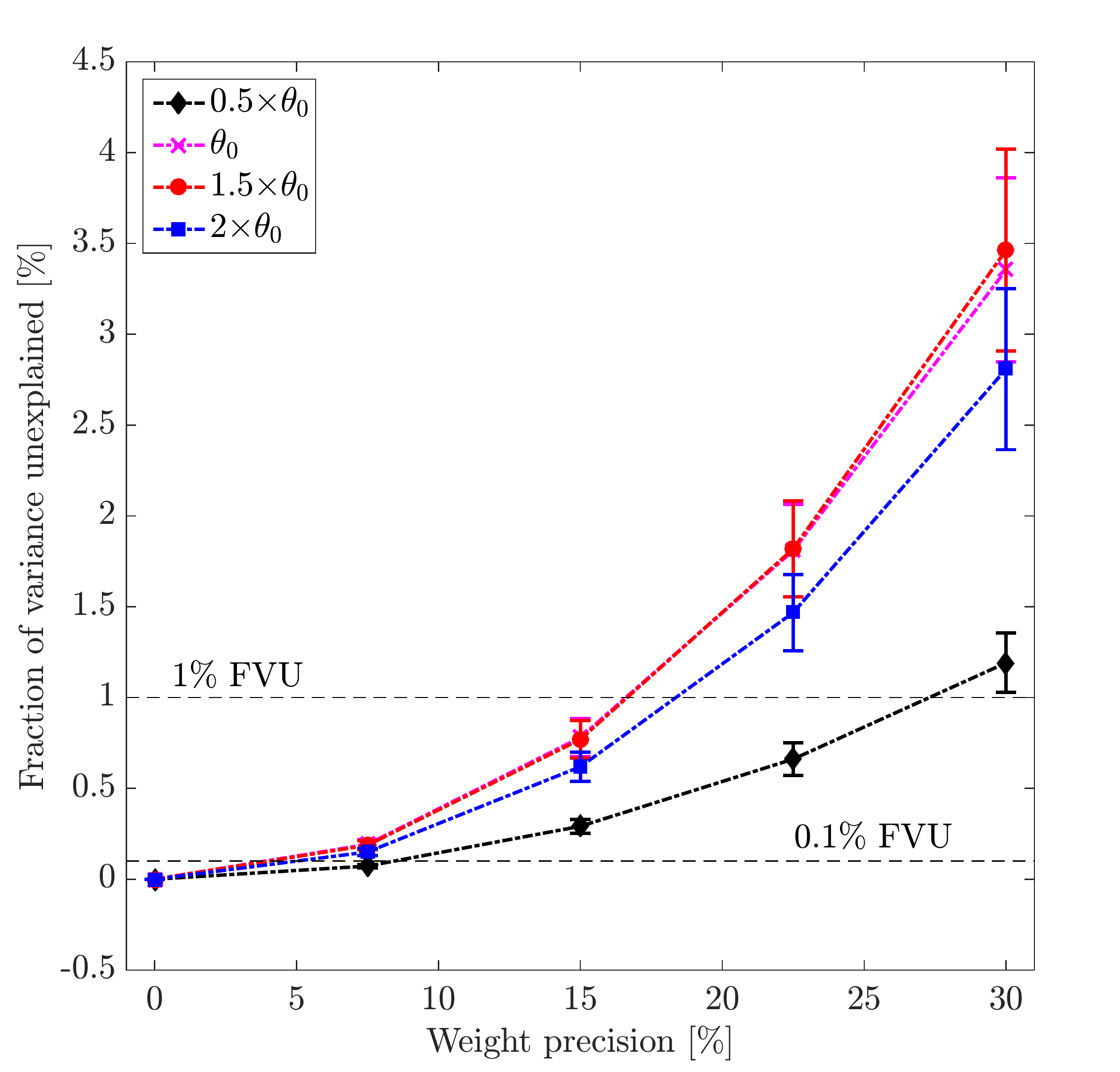}
	\captionof{figure}{\small  FVU as function of \textbf{Top:} height accuracy  \textbf{Bottom:} weight accuracy.}
	\label{F:fvuVweight}
\end{figure}

\begin{figure}
	\centering
	\includegraphics[height=8cm]{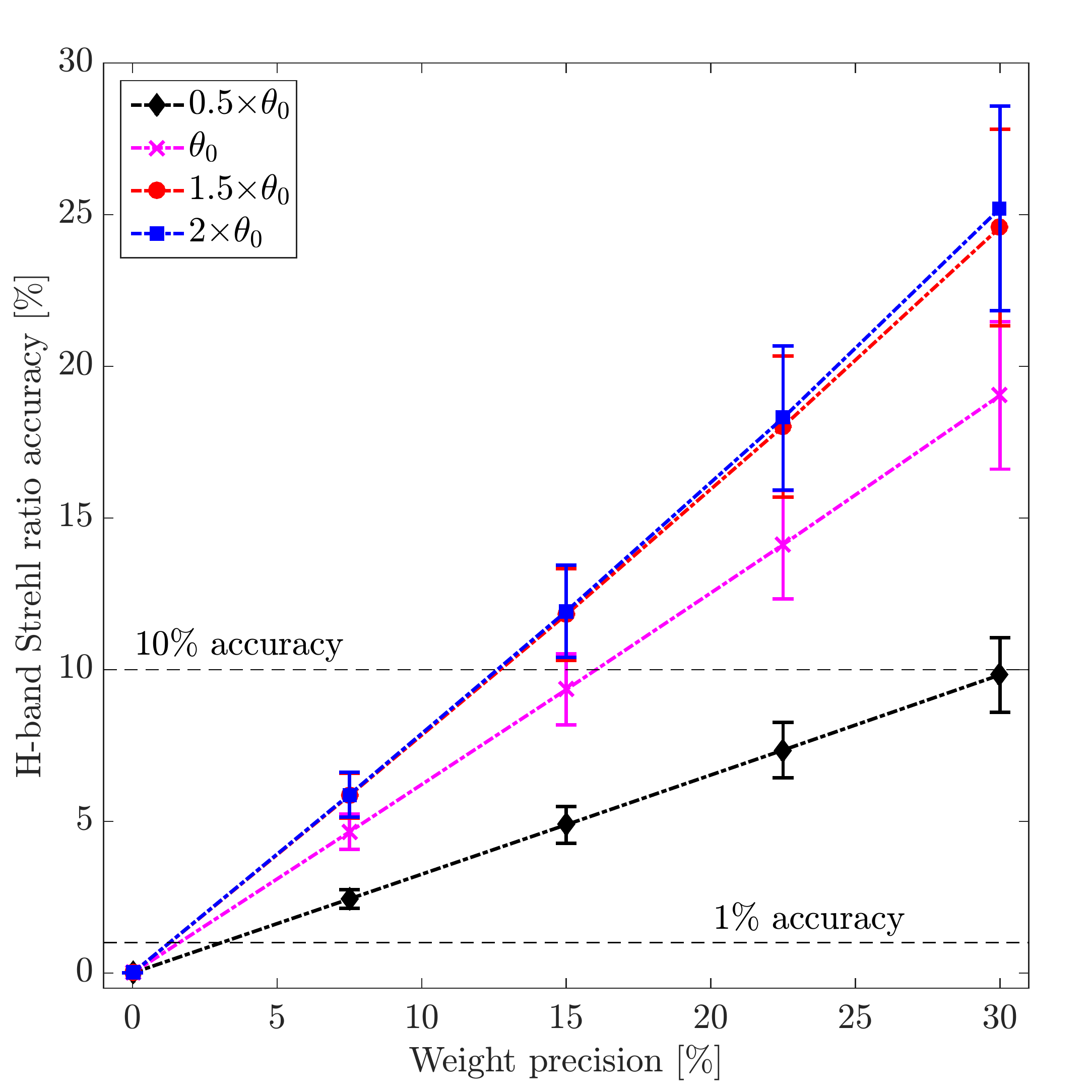}
	\includegraphics[height=8cm]{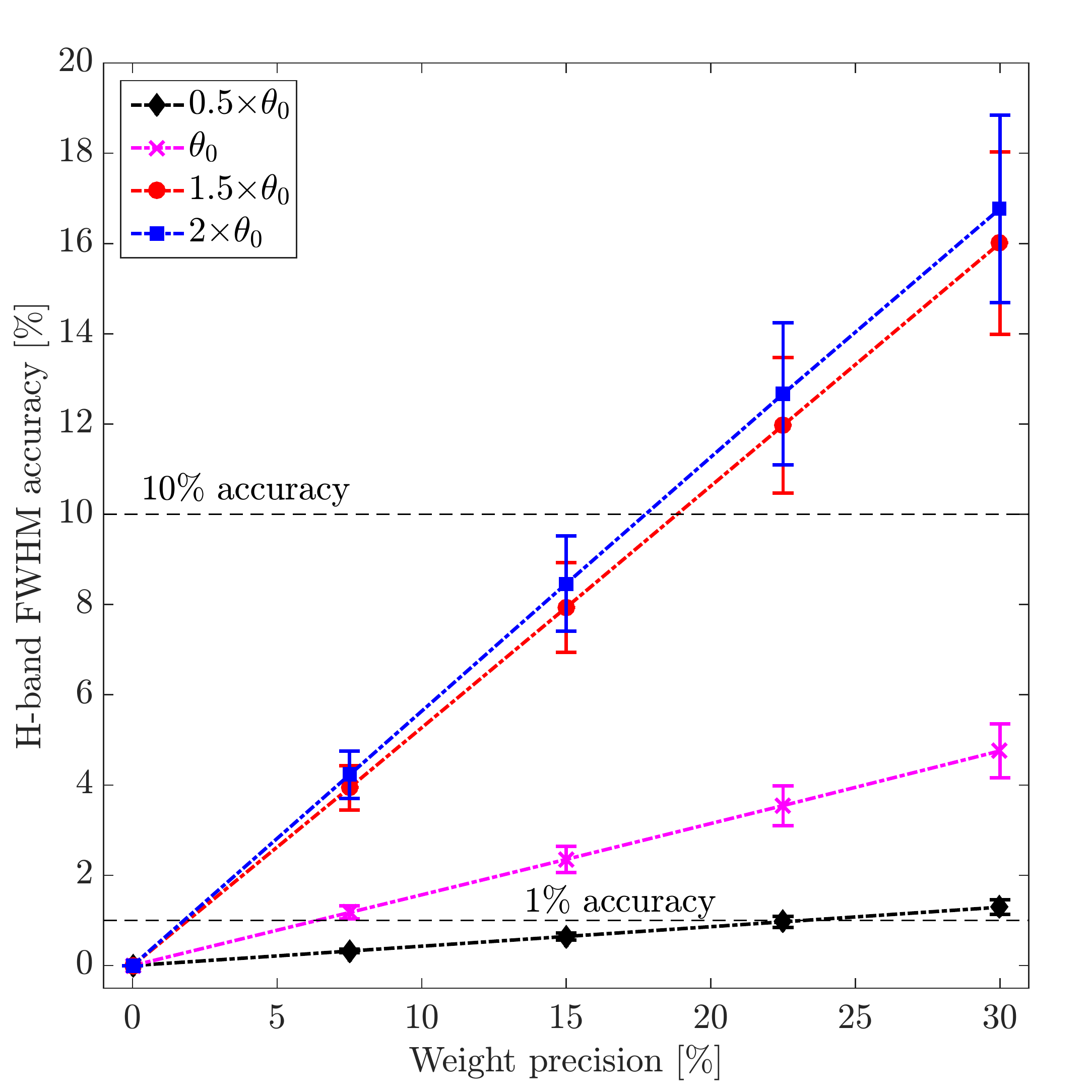}
	\captionof{figure}{\small  \textbf{Top:} Strehl ratio  \textbf{Bottom:} FWHM  accuracy versus accuracy on weights estimation in H-band.}
	\label{F:srfwhmVweight}
\end{figure}

\begin{figure}
	\centering
	\includegraphics[height=8cm]{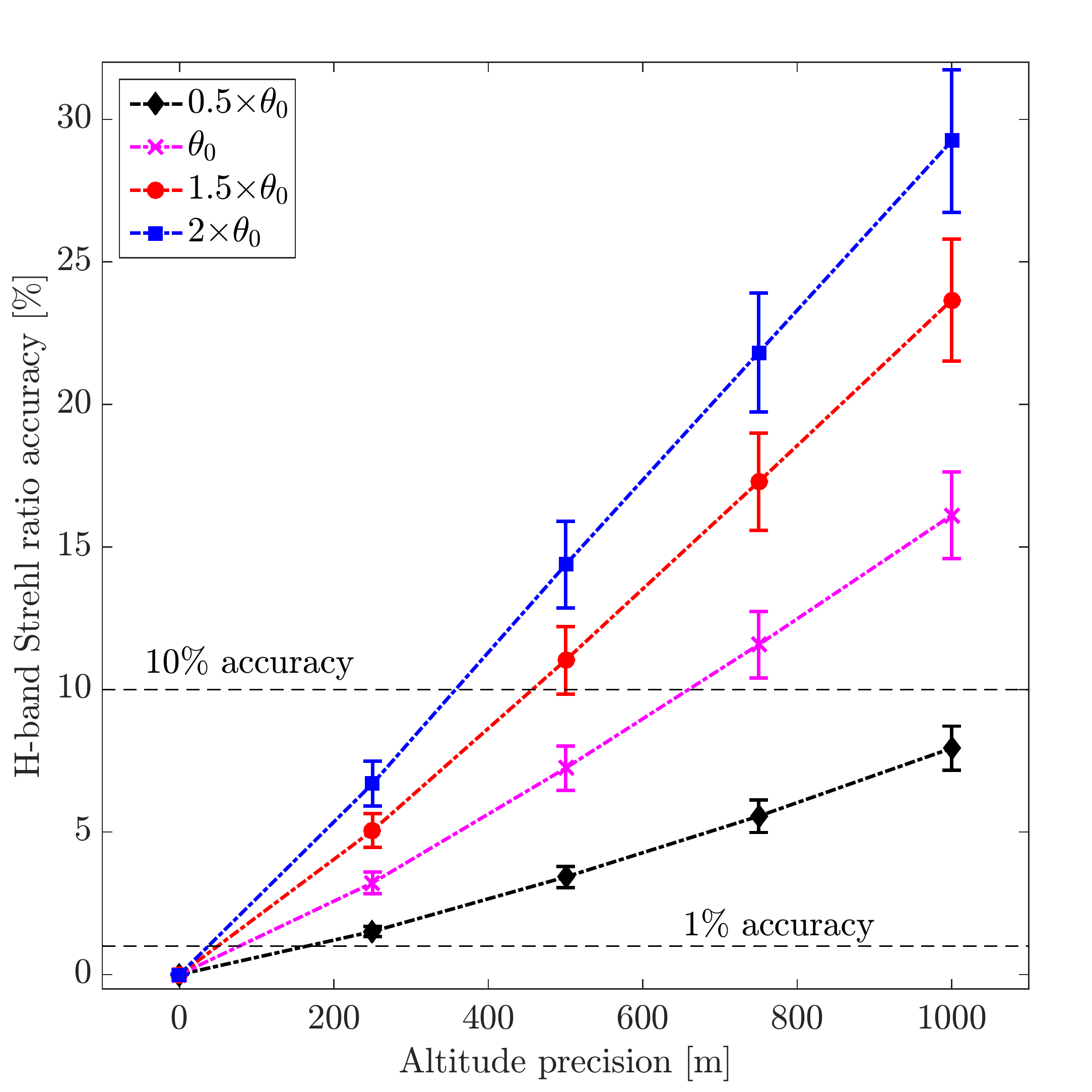}
	\includegraphics[height=8cm]{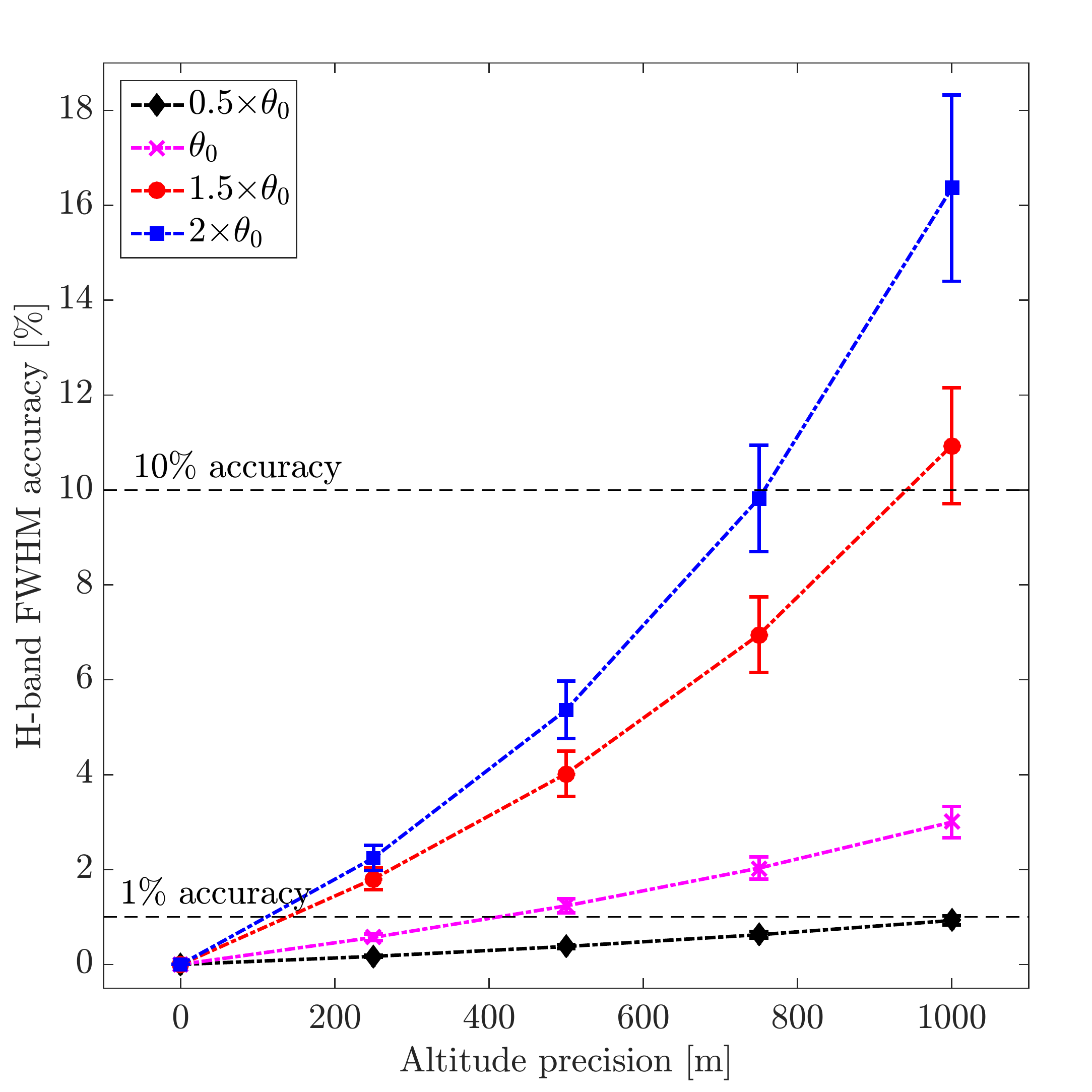}
	\captionof{figure}{\small \textbf{Top:} Strehl ratio \textbf{Bottom:} FWHM accuracy versus precision on heights estimation in H-band.}
	\label{F:srfwhmVheight}
\end{figure}

\begin{figure}
	\centering
	\includegraphics[height=7.75cm]{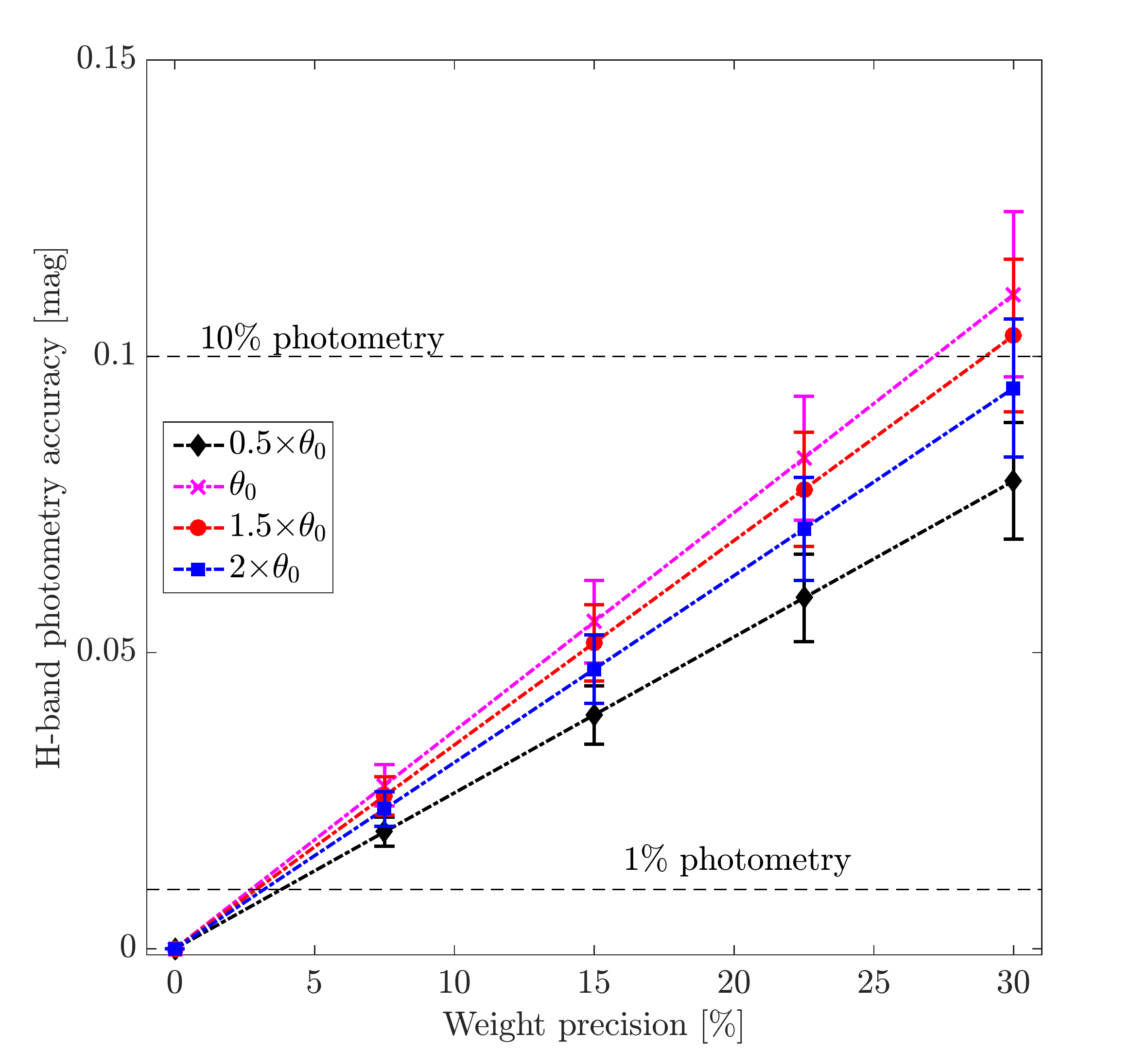}
	\includegraphics[height=8cm]{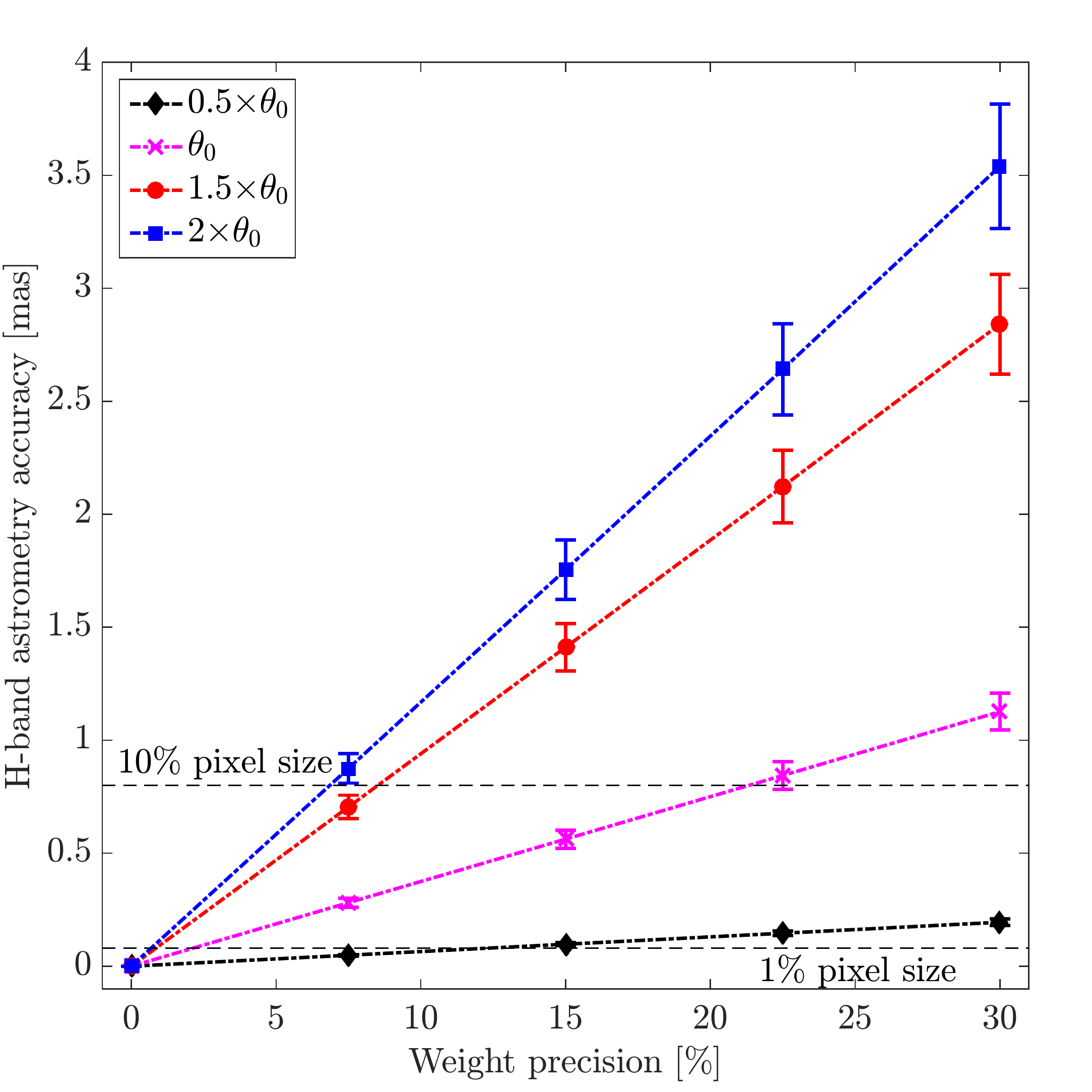}
	\captionof{figure}{\small \textbf{Top:} Photometry \textbf{Bottom:} astrometry accuracy versus precision on weights estimation in H-band.}
	\label{F:photoastroVweight}
\end{figure}

\begin{figure}
	\centering
	\includegraphics[height=8.75cm]{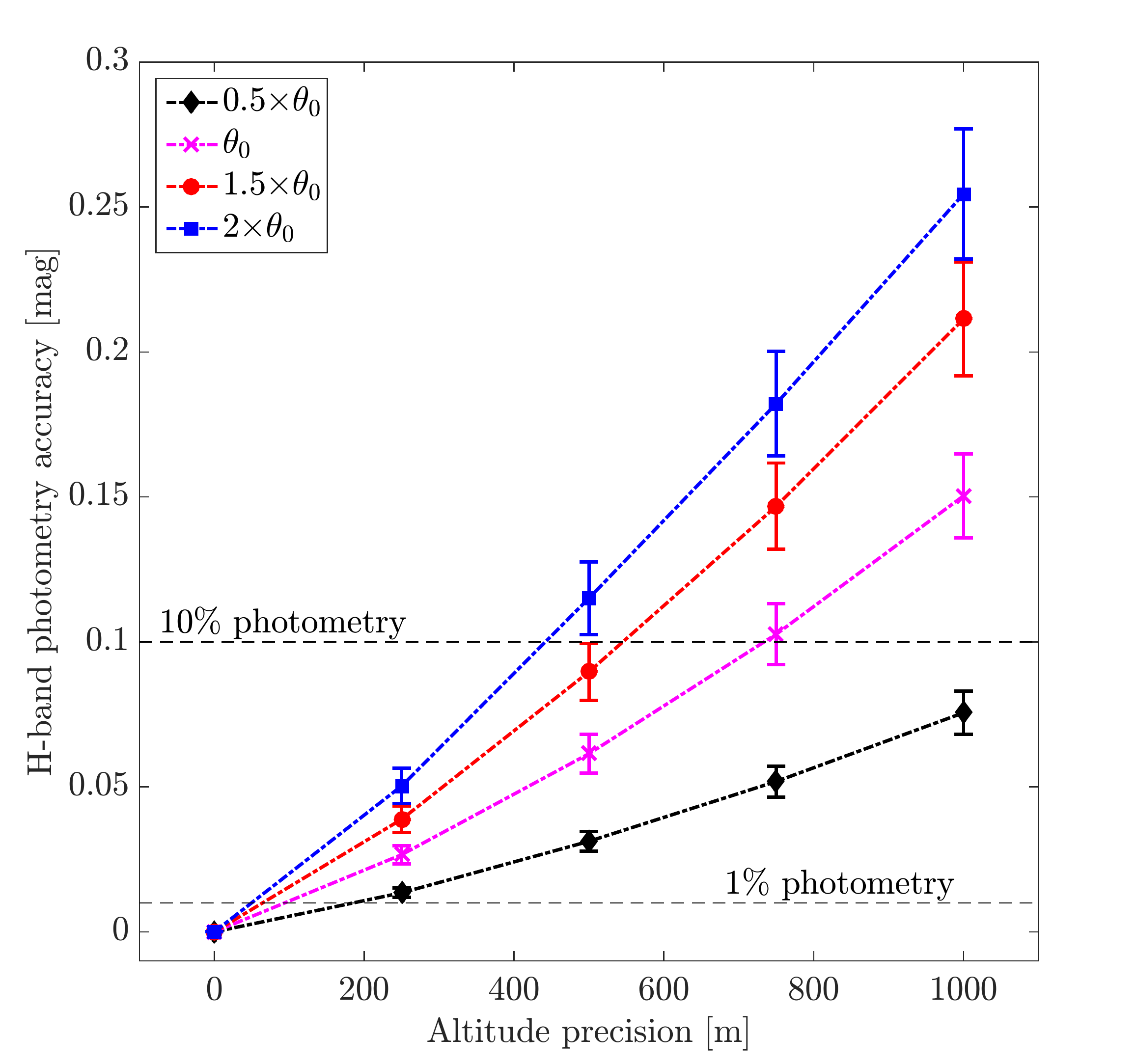}
	\includegraphics[height=9cm]{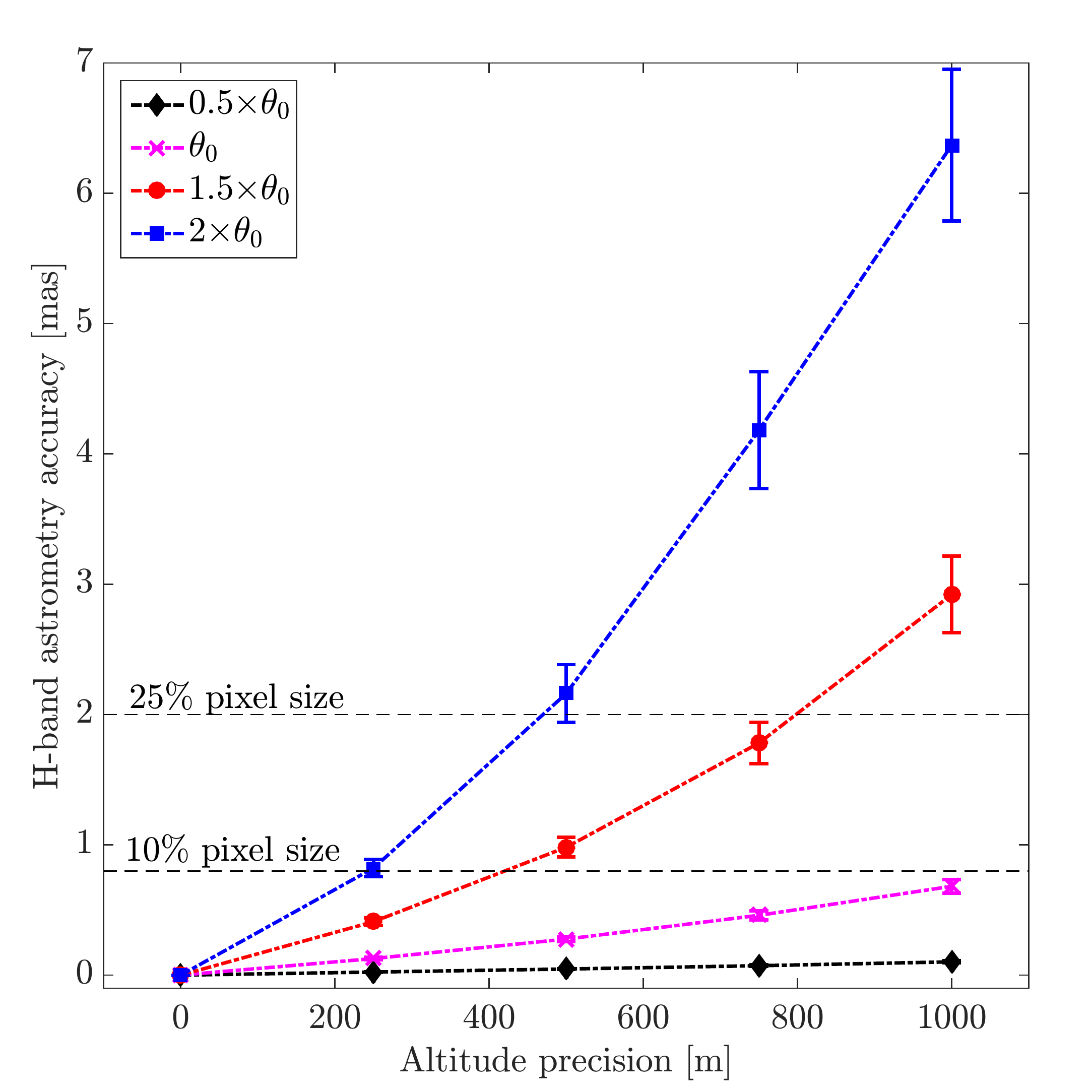}
	\captionof{figure}{\small \textbf{Top:} Photometry \textbf{Bottom:} astrometry accuracy versus precision on heights estimation in H-band.}
	\label{F:photoastroVheight}
\end{figure}

Figs.~\ref{F:srfwhmVweight},~\ref{F:photoastroVweight},~\ref{F:srfwhmVheight} and~\ref{F:photoastroVheight} provide curves of mean values at multiple separations versus weights and heights precision given by $\sigma_p$. Also Fig.~\ref{F:fvuVweight} illustrates FVU errors regarding accuracy on height and weight. Curves envelopes are not represented, but range as similarly as they do in Figs.~\ref{F:fvuVnlayer},~\ref{F:srfwhmVnLayer} and~\ref{F:photoastroVnLayer} for a 7-layers profile.

Globally, PSF metrics deviate monotonically with respect to inputs precision level, with a speed that grows with the angular separation from the guide star. It is an expected results: the PSF model differs more largely when introducing more errors on inputs and for stronger anisoplanatism cases.
Strehl ratio is known to follow a $\exp(\theta_0^{-5/3})$ law while FWHM is proportional to $\theta_0^{-5/3}$. This latter is given by $r0.(\sum(h^{5/3}.\cnh)^{3/5}$ that makes FWHM proportional to $\rz^{-5/3}.\sum(h^{5/3}.\cnh)$ and SR proportional to $1+ (r0^{-5/3}.\sum(h^{5/3}.\cnh))$ for small amount of variations on inputs. Weights as introduced in Fig.~\ref{F:fvuVweight} scale with $r0^{-5/3}$; we directly see why FWHM/SR are supposed to be linear regarding the weight. About the altitude, both FWHM and SR should not be linear in h, but because the exponential term involved in the Strehl expression, the non-linear regime on the Strehl appears after 1km of altitude precision, while we do not have this mitigation effect of the non-linear component on the FWHM that follows directly a 5/3 power law on altitude.

Strehl and FWHM are estimated within 1\% accuracy as long as we ensure a precision of 10\% on weights and roughly 200~m on heights. This latter means that heights of the retrieved 7-layers profile must be measured within 200~m precision at 1 $\sigma$. Astrometry is more critical regarding the weight precision that must be ensured to a 7\%-level to get an astrometry of 10\% of the pixel scale. Also, photometry is only impacted by height precision that also must provided within 200~m to get 1\% level of photometry. Such a level of accuracy on $\cnh$ weights is accessible from external profilers~\citep{Butterley2006}, but altitude resolution reaches generally 500~m up to 1~km, leading to an accuracy of 5 up to 10\% on PSF estimates, which can be still acceptable for some science cases that are noise-limited for instance.
Fig.~\ref{F:srfwhmVweight} highlights the error introduced on weights does not degrade the PSF linearly with the separation. Errors at $2\times\theta_0$ are lower compared to $1.5\times\theta_0$. There is a physical explanation: at such a separation, the phase is largely decorrelated. Eq.\ref{E:covDelta_zonal} involves both the phase auto-covariance and cross-covariance terms. The latter converge towards zero when the separation goes to infinity. However these are the terms that carry the sensitivity to the fractional weights of the $\cnh$ profile, which means the off-axis PSF is less and less sensitive to weights precision for increasing separations.

How do our results translate to ELT-scales? Anisoplanatism variance degrades with $(\theta/\theta_0)^{5/3}$, where $\theta_0$ is only an atmosphere-dependent variable; it therefore seems to us the findings ought to be conserved at an ELT scale, in a way we would need 10ish layers to describe the anisoplanatism. However our analysis is focused on the anisoplanatism only and does not include any tomography or multi-laser configuration. On top of that, others metrics may be more relevant regarding the science case. As a conclusion, we have an order of magnitude for the number of layers of about 10 for anisoplanatism characterisation only.

\subsection{Metrics correlation}

\begin{figure}
	\centering
	\includegraphics[height=8.75cm]{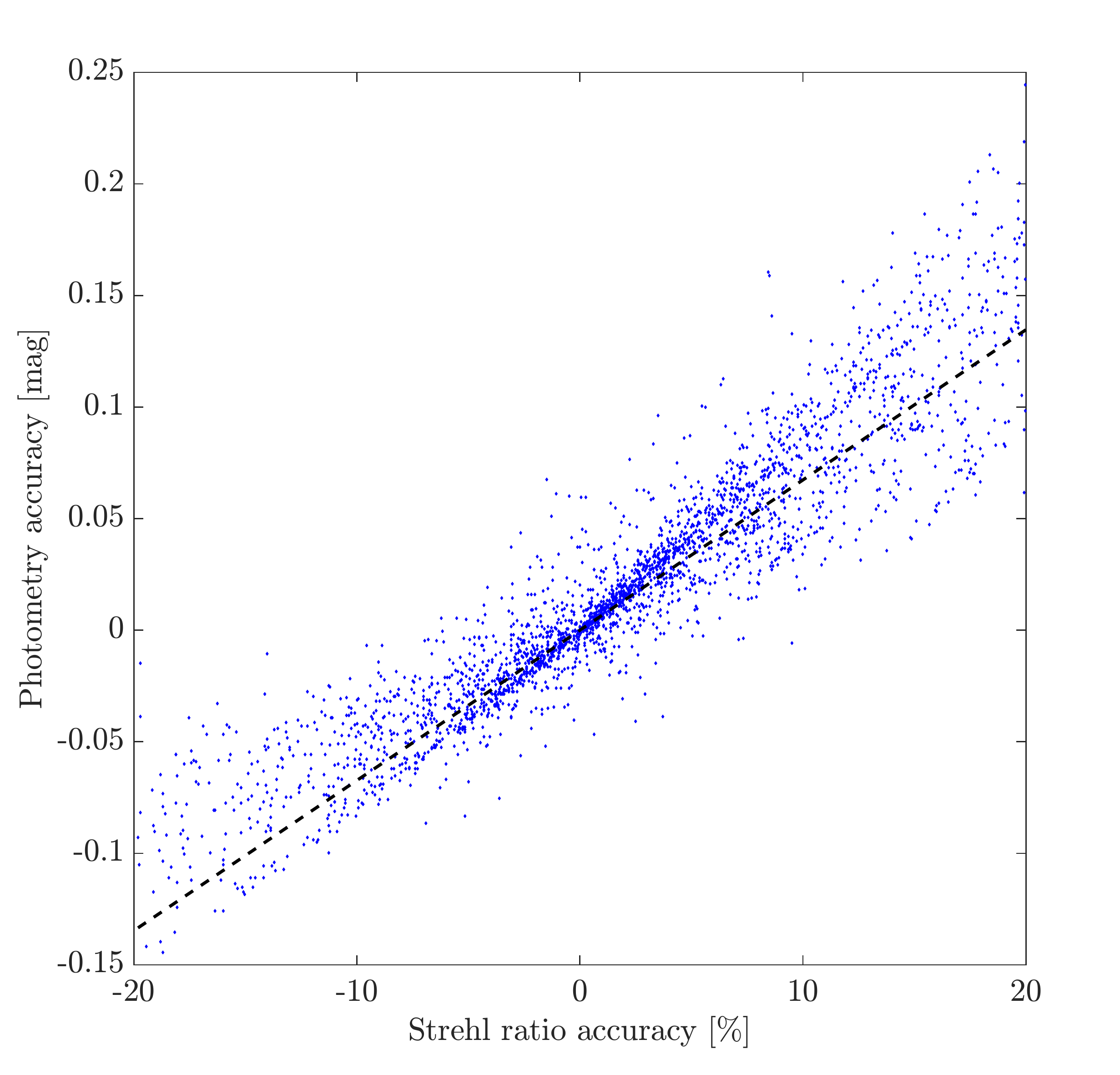}
	\includegraphics[height=8.75cm]{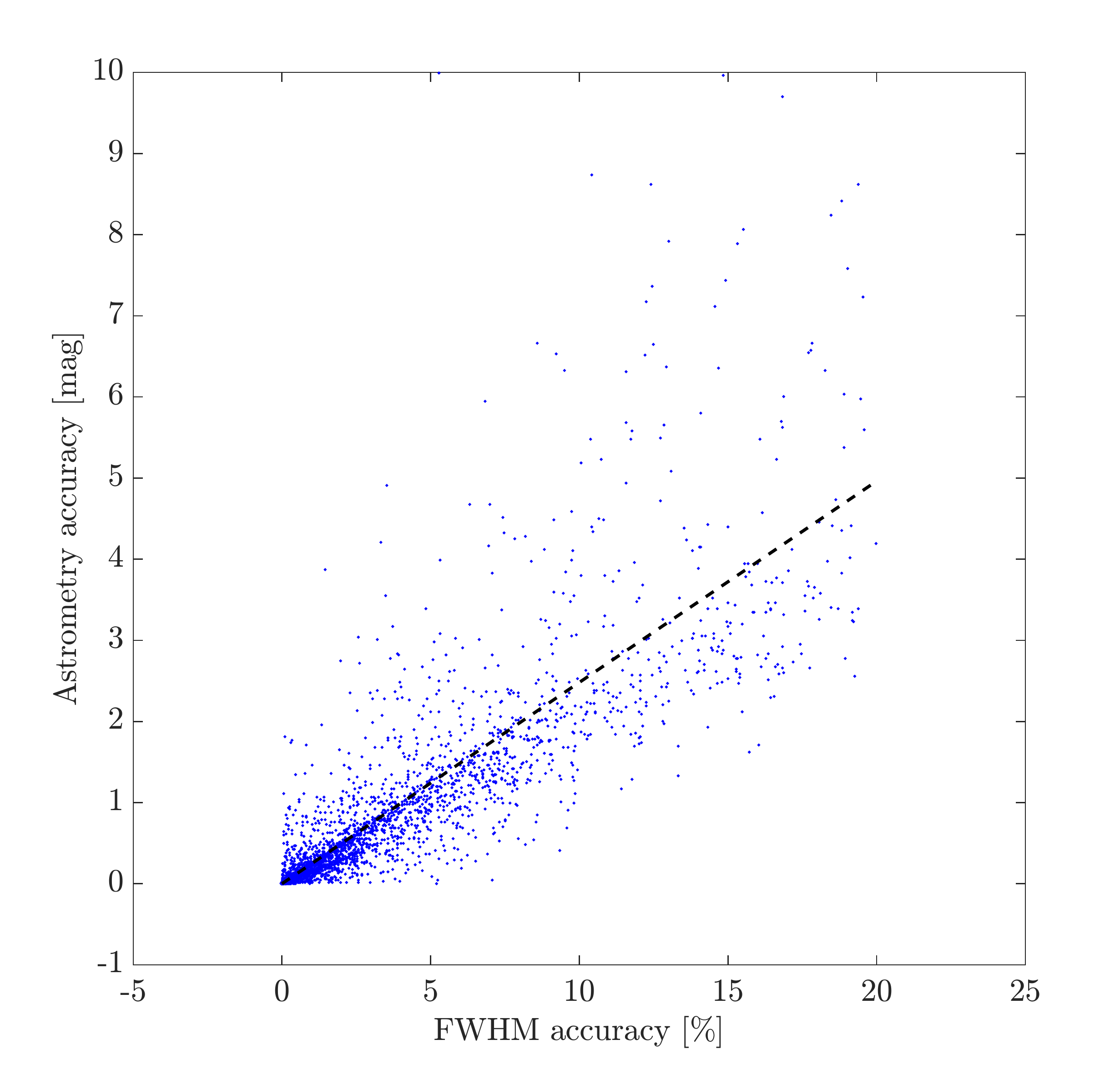}
	\caption{\textbf{Up:} Photometry accuracy versus Strehl ratio accuracy \textbf{Down:} Astrometry accuracy versus FWHM accuracy}
	\label{F:correl}
\end{figure}

We have gathered all metrics values from previous analyses and compared to each other in order to track correlations. We ended with 
the following regression relationship:
\begin{equation}
\begin{aligned}
	&\Delta \text{mag [mag]}  = 0.0067 \pm 0.00052 \times \Delta \text{SR [\%]}\\
	&\Delta \text{ast [mas]} = 0.25 \pm 0.024 \times \Delta \text{FWHM [\%]}
\end{aligned}	
\end{equation}
that allows to match accurately observations as illustrated in Fig.~\ref{F:correl}.
We have noticed a quadratic dependency of the FVU with the photometry~(and so Strehl-ratio as well) accuracy; for accurate photometric measurements~($\Delta$mag< 5\%), we have FVU$~=5.4\pm 0.6\times\Delta$mag,while for less accurate measurements, we get FVU$~=24\pm 3\times\Delta$mag. We also observed a clear correlation trend between FVU and astrometry accuracy values, given by  FVU$~=1.2\pm 0.2\times\Delta$ast, although we observed a large discrepancy of samples around the linear regression. In summary, FVU is an efficient metric to characterize the PSF; it defines a comprehensive scale value that depends on both PSF-related parameters and key science observables.

\section{Towards ELTs~: sensitivity analysis validated using HeNOS}
\label{S:HENOS}

HeNOS (Herzberg NFIRAOS Optical Simulator) is a MCAO test bench designed to be a scaled down version of NFIRAOS,
the first light adaptive optics system for the Thirty Meter Telescope~\citep{Mieda2018,Rosensteiner2016}. We used HeNOS in single-conjugated mode with in closing the loop on ones of the 4 LGSs distributed over a 4.5"-side length square constellation, while atmosphere is created using three phase screens. Summary of main parameters is given in Tab.~\ref{T:setup}. To simulate the expected PSF degradation across the field on NFIRAOS at TMT, all altitudes are stretched up by a factor 11. Moreover, at the time we acquired HeNOS data, science camera was conjugated at the LGSs altitude; LGSs beams are propagating along a cone but are arriving in-focus at the science camera entrance.
\begin{table}
	\centering
	\begin{tabular}{l|l}
		Asterism side length  & 4.5"\\
		Sources wavelength    & 670 nm \\
		$r_0$ (670 nm)        & 0.751  \\      
		$\theta_0$ (670 nm)	  & 0.854"  \\                   
		fractional $r_0$      & 74.3\% ,17.4\%,8.2\% \\
		altitude layer        & (0.6, 5.2, 16.3) km\\
		source height         & 98.5 km \\
		Telescope diameter    & 8.13~m\\
		DM actuator pitch     & 0.813~m\\
	\end{tabular}
	\caption{HeNOS set up summary.}
	\label{T:setup}
\end{table}

We did acquire closed-loop data when guiding on a LGS in single-conjugated mode in July 2017. Two data sets have been acquired respectively with and without phase screens in the LGSs beams. This second measurement allows to measure the best Non common path aberrations~(NCPA)-limited PSFs over all directions. See~\citep{Lamb2016} about phase diversity/focal plane sharpening methods deployed on HeNOS to calibrate NCPA.

Our purpose is to demonstrate the anisoplanatism model we developed allows for a good representation of PSF characteristics within the expected accuracy given by the sensitivity analysis. We derive off-axis PSF from the PSF in the guide-star direction by the use of Eq.~\ref{E:otf2t1}. Although off-axis PSFs are largely dominated by anisoplanatism, static aberrations that vary across the field must be taken into account to improve the PSF modeling.

We have compensated for on-axis NCPA and put back off-axis static aberrations for each individual direction by convolving PSFs with best NCPA-limited PSF. These PSF includes both NCPA calibration and field static aberrations residual in off-axis direction. OTF in the off-axis direction $\theta$ is thus yielded by the following calculation
\begin{equation} \label{E:otfHenos}
	\otf{}(\theta) = \otf{}(0) \times \dfrac{\otf{\text{stat}}(\theta)}{\otf{\text{stat}}(0)} \times\text{ATF}_\Delta(\theta)
\end{equation}
where $\otf{\text{stat}}(\theta)$ is derived from observations without phase screens in direction $\theta$ while $\otf{}(0)$ is the on-axis OTF during AO operation on phase screens.

\begin{figure*}
	\centering
	\includegraphics[width=12cm]{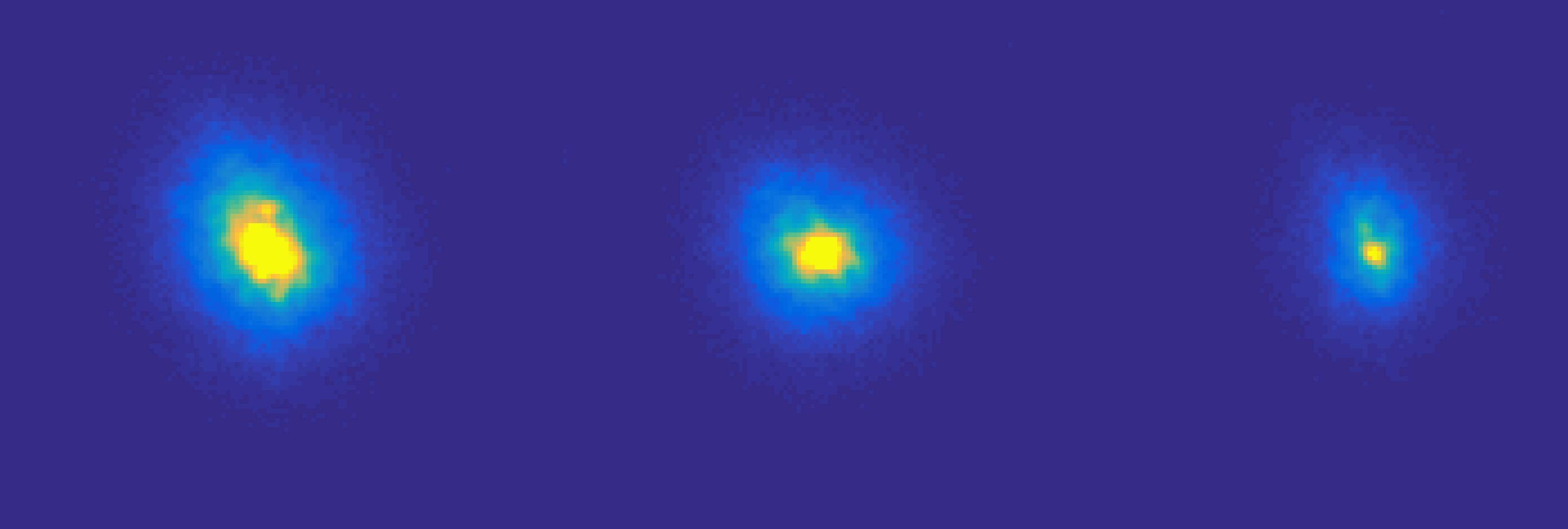}\\
	\includegraphics[width=12cm]{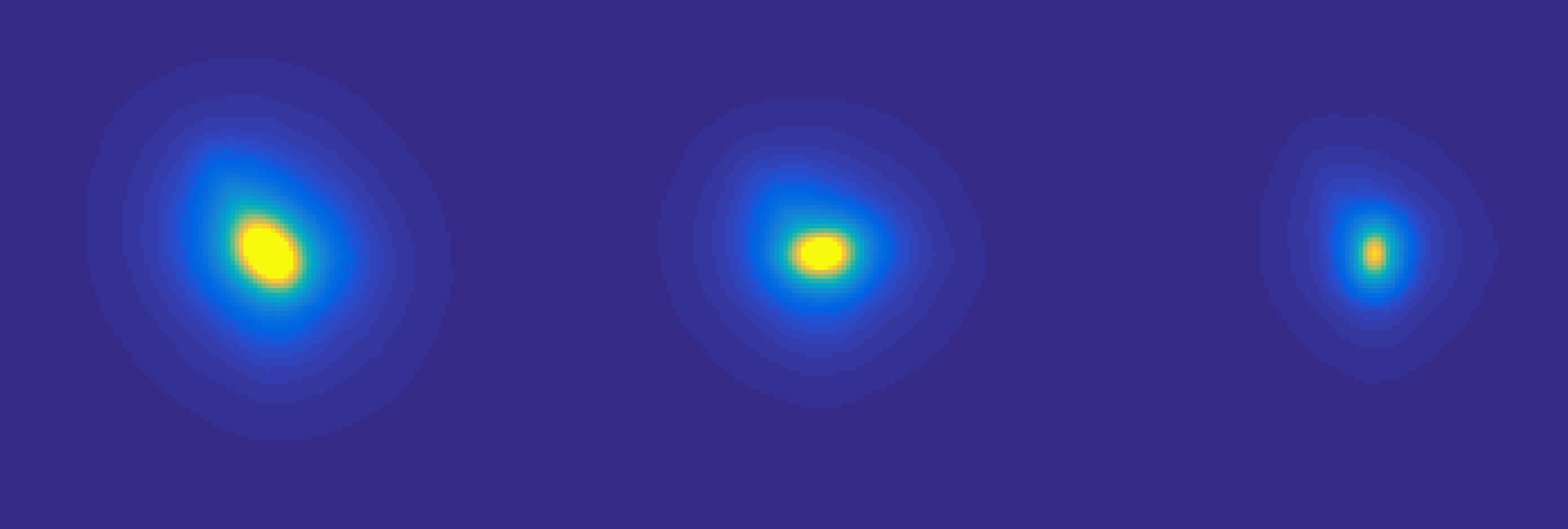}\\
	\hspace{.1cm}\includegraphics[width=12cm]{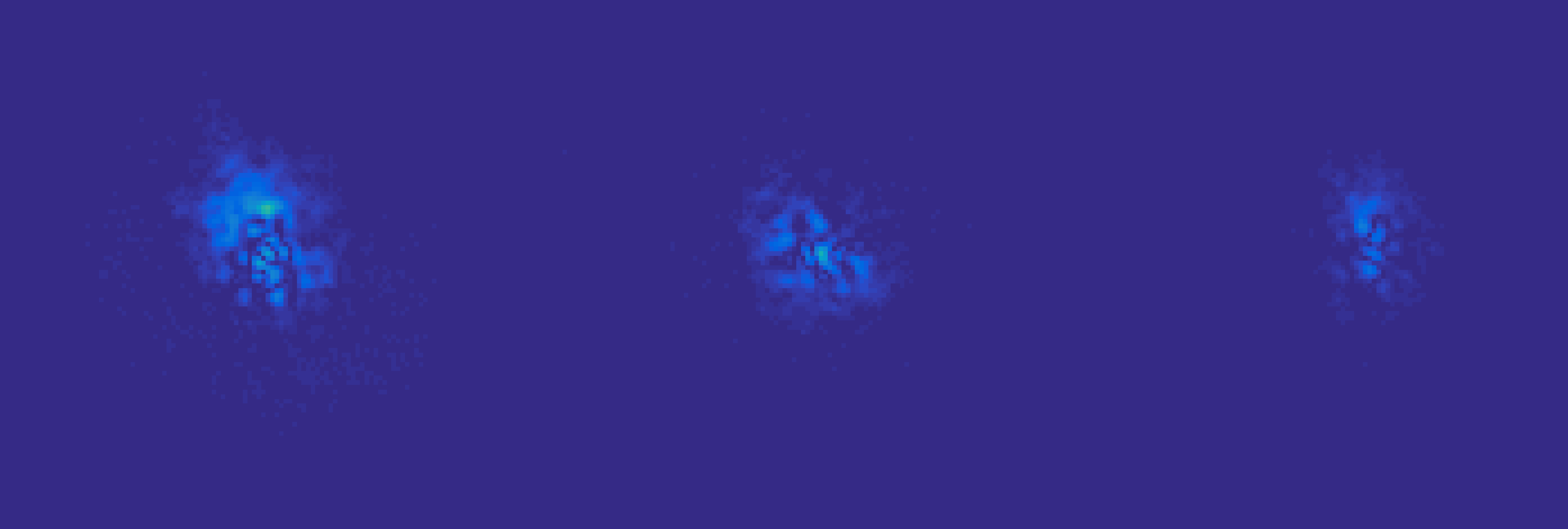}
	\captionof{figure}{\textbf{Up~:} 670 nm HeNOS PSFs while running the AO loop on LGS 1 \textbf{Middle~:} PSFs predicted from on-axis PSF and anisoplanatism \textbf{Down~:} Residual on the prediction.}
	\label{F:henosPSFs}
\end{figure*}

Because all sources are focused at the same height, we only need to consider angular anisoplanatism to extrapolate the on-axis PSF to any other direction. Fig.~\ref{F:henosPSFs} provides focal-plane images acquired in visible~(670 nm) whilst closing the AO loop on LGS 1 (top-left PSF). Anisoplanatism on LGS 2 (bottom-right), 3 (bottom left) and 4 (top right) is clearly visible and produces a strong PSF elongation in the guide star direction. We also illustrate off-axis PSFs modelled derived from Eq.~\ref{E:otfHenos}. At a glance we are capable of reproducing the good shape and elongation of off-axis PSFs. We report in Tab.~\ref{T:henosStats} a quantification of PSF characteristics measured versus estimated, that shows that we get 10\%-level of accuracy on PSF metrics and reach 5\% of fraction of variance unexplained on all PSFs.

\begin{table}
	\centering
	\small
	\begin{tabular}{|c|c|c|c|c|c|c|}
		\hline
		& \multicolumn{3}{c|}{Bench} &\multicolumn{3}{c|}{Predicted}\\
		\hline
		\# LGS & 2 & 3 & 4 &2 &3 &4\\
		\hline
		SR[\%] & 4.1& 6.0&5.0 &3.2&4.2 &4.2\\
		\hline
		FWHM [mas] &117 & 90 &100 & 105&84 &82\\
		\hline
		Aspect ratio & 1.41& 1.27 & 1.42& 1.41&1.34 &1.33\\
		\hline
		EE [\%] & 53& 56 & 64&65 &67 &64\\
		\hline
		FVU$_\text{PSF}$ [\%] & x & x  & x & 5.8 & 4.1 & 6.0\\
		\hline
	\end{tabular}
	\caption{PSF characteristics measured on laboratory PSF compared to predict values from on-axis PSF and anisoplanatism. EE is taken at $10\lambda/D$ and FVU values are estimated using Eq.~\ref{E:FVU} is derived using 60 pixels.}
	\label{T:henosStats}
\end{table}

Residuals are mostly composed by a combination of high spatial frequencies patterns and a large structure oriented towards the guide star direction. The first component may be introduced by residual speckles and static aberrations, while the second feature suggests the anisoplanatism effect is not perfectly well characterized. It echoes the previous discussion on sensitivity: the parameters in Tab.~\ref{T:setup} we have considered as inputs of our model have been measured with a certain accuracy. Weights are estimated within 10\% while heights are retrieved every 1~km. 

\begin{figure}
	\centering
	\includegraphics[height=8cm]{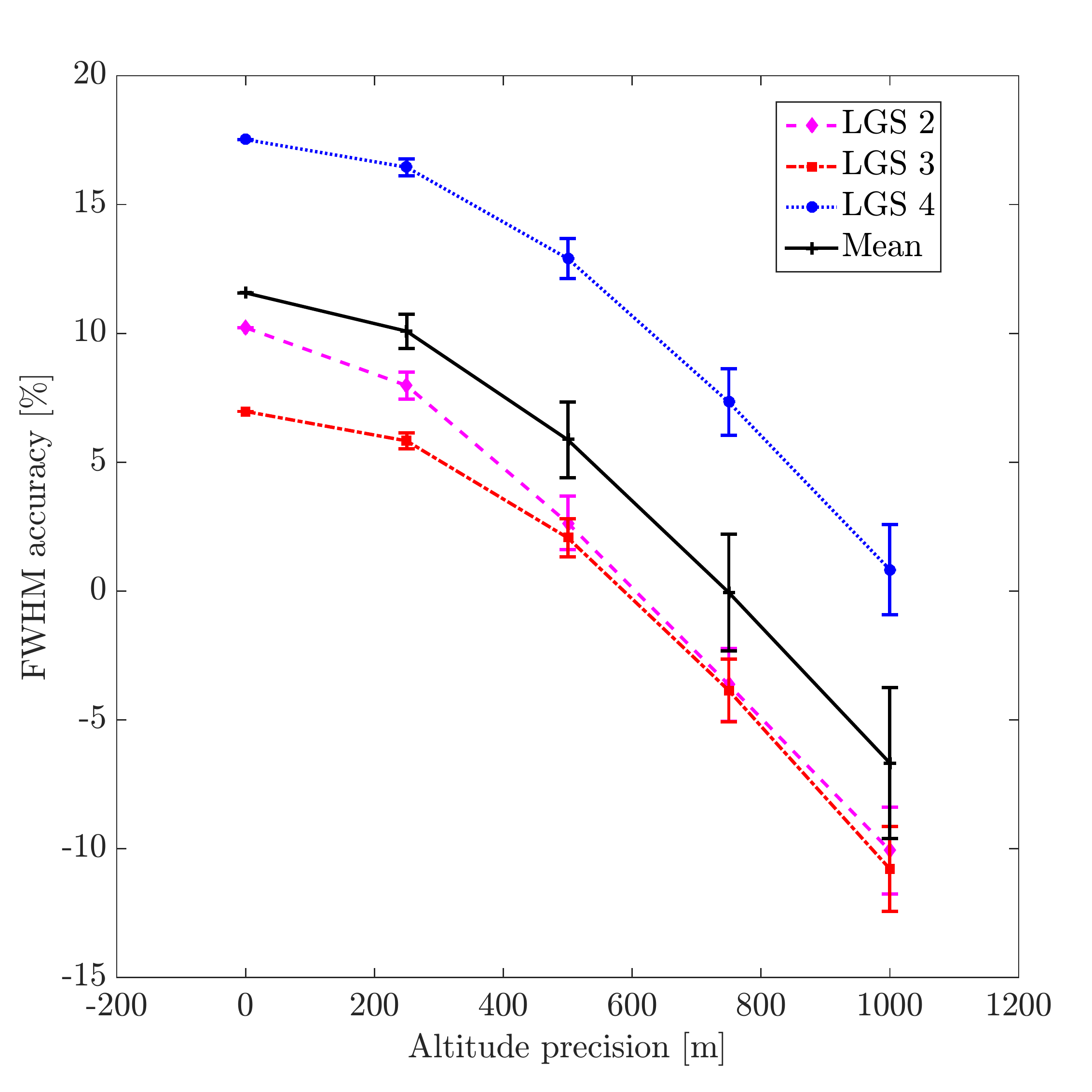}
	\caption{\small PSF FWHM predicted versus the absolute precision in altitude layers. Plots represent FWHM variations on each individual off-axis PSF as the overall mean FWHM over PSFs as well.}
	\label{F:FWHMhenosvdH}
\end{figure}

To confirm whether such a level of accuracy may explain bias on PSF estimates, we have deliberately introduced a random error on $\cnh$ following the methodology presented in Sect.~\ref{S:sensitivity}. We illustrate in Fig.~\ref{F:FWHMhenosvdH} how the off-axis PSF FWHM vary regarding the layer height precision. At this range of separations~(4.5" compared to 0.854" for $\theta_0$), the anisoplanatism characterization does not depend on the weight precision as discussed earlier and it has been confirmed for HeNOS. In Fig.~\ref{F:FWHMhenosvdH}, we highlight that the relative error on FWHM, averaged out the three LGS, reaches zero for an altitude precision of 750~m. From~\citep{Rosensteiner2016}, altitude have been measured at every 1~km, which comply with our results. It does not mean we will get a better PSF characterization by shifting all layers by this quantity; instead it confirms the precision on inputs parameters translate into accuracy on PSF-metrics. In next works, we will particularly investigate for inverting the problem, in order to retrieve the $\cnh$ profile from a collection of observed off-axis PSFs.

\section{Conclusions}
\label{S:conclusions}

We carried out developments to characterise anisoplanatism in LGS (and NGS) systems in order to estimate off-axis PSFs using a point-wise method that is accurate, numerically efficient and can potentially be used with 40~m class telescopes featuring a large number of degrees of freedom. For a 7-layers median profile at Paranal, we have demonstrated our modelling complies with physical-optics simulations to within 0.1\% on the fraction of variance unexplained, together with 1\% on PSF-related metrics, as Strehl ratio and FWHM, with a LGS and a NGS at respectively 20" and 40" off-axis. Stars flux and position are conserved respectively at a mill-mag and $\mu$arcsec-level. Finally, the total anisoplanatism with LGS is accurately split into a focal+angular and a tilt component.

With the purpose of determining how $\cnh$ knowledge constrains off-axis PSFs, we have investigated how input parameter errors on the number of layers, its heights and weights translate into PSF errors. Firstly, we have highlighted for 10~m class telescopes that percent-level accuracy on SR, FWHM and FVU can be met with seven turbulent layers. On tight binaries, photometry can be retrieved at 3\% while astrometry stays within 5\% of pixel size. If we may admit 10\% of accuracy on PSF metrics, $\cnh$ must be provided within 500 m accuracy on altitude height. Those results can be extended to ELT scale, with a conclusion that we need roughly ten layers to characterize at 1\%-level a PSF on a 40~m class telescope.

We have finally validated this methodology on the HeNOS testbed where $\cnh$ values and its accuracy are known; By introducing random errors on anisoplanatism model input parameters, we have improved the off-axis PSF FWHM estimation. It demonstrates the precision in those parameters measurements by explaining the bias we have observed on PSF off-axis PSF modelling.

Next works will aim at applying our analytical formulation to real on-sky observations of crowded fields at Keck for improving the PSF characterisation and the estimation of key science parameters. Moreover we are also focusing in inverting the problem in order to retrieve the $\cnh$ profile from a collection of observed off-axis PSFs. Such a technique is investigated for providing an accurate reconstructed PSF for single-conjugated AO systems for which $\cnh$ profile is not measurable from the AO control loop data.

\section*{Acknowledgments}
This work was supported by the A*MIDEX project (no. ANR-11-IDEX-0001-02) funded by the "Investissements d'Avenir" French Government programme, managed by the French National Research Agency (ANR).

\bibliographystyle{plain} 
\bibliography{biblioLolo}

\begin{thebibliography}{10}

\bibitem{Ascenso2015}
J.~{Ascenso}, B.~{Neichel}, M.~{Silva}, T.~{Fusco}, and P.~{Garcia}.
\newblock {PSF reconstruction for AO photometry and astrometry}.
\newblock In {\em Adaptive Optics for Extremely Large Telescopes IV (AO4ELT4)},
  December 2015.

\bibitem{Bertin1996}
E.~{Bertin} and S.~{Arnouts}.
\newblock {SExtractor: Software for source extraction.}
\newblock {\em \aap}, 117:393--404, June 1996.

\bibitem{Britton2006}
M.~C. {Britton}.
\newblock {The Anisoplanatic Point-Spread Function in Adaptive Optics}.
\newblock {\em \pasp}, 118:885--900, June 2006.

\bibitem{Butterley2006}
T.~{Butterley}, R.~W. {Wilson}, and M.~{Sarazin}.
\newblock {Determination of the profile of atmospheric optical turbulence
  strength from SLODAR data}.
\newblock {\em \mnras}, 369:835--845, June 2006.

\bibitem{Chassat1989}
F.~{Chassat}.
\newblock {Calcul du domaine d'isoplan{\'e}tisme d'un syst{\`e}me d'optique
  adaptative fonctionnant {\`a} travers la turbulence atmosph{\'e}rique.}
\newblock {\em Journal of Optics}, 20:13--23, February 1989.

\bibitem{Conan1994}
{J.-M.} {Conan}.
\newblock {\em {\'E}tude de la correction partielle en optique adaptative}.
\newblock PhD thesis, Universit{\'e} Paris XI Orsay, October 1994.

\bibitem{Conan2010}
R.~{Conan}, C.~{Bradley}, O.~{Lardi{\`e}re}, C.~{Blain}, K.~{Venn},
  D.~{Andersen}, L.~{Simard}, J.-P. {V{\'e}ran}, G.~{Herriot}, D.~{Loop},
  T.~{Usuda}, S.~{Oya}, Y.~{Hayano}, H.~{Terada}, and M.~{Akiyama}.
\newblock {\em {Raven: a harbinger of multi-object adaptive optics-based
  instruments at the Subaru Telescope}}, volume 7736 of {\em \procspie}.
\newblock July 2010.

\bibitem{Conan2014OOMAO}
R.~{Conan} and C.~{Correia}.
\newblock {Object-oriented Matlab adaptive optics toolbox}.
\newblock In {\em Adaptive Optics Systems IV}, volume 9148 of {\em \procspie},
  page 91486C, August 2014.

\bibitem{Correia2011}
C.~{Correia}, J.-P. {V{\'e}ran}, B.~{Ellerbroek}, L.~{Gilles}, and L.~{Wang}.
\newblock {Laser-Guide Star Point-Spread Function Reconstruction for ELTs}.
\newblock In {\em Second International Conference on Adaptive Optics for
  Extremely Large Telescopes. Online, id.70}, page~70, September 2011.

\bibitem{Diolaiti2000}
E.~{Diolaiti}, O.~{Bendinelli}, D.~{Bonaccini}, L.~M. {Close}, D.~G. {Currie},
  and G.~{Parmeggiani}.
\newblock {StarFinder: A code for stellar field analysis}.
\newblock Astrophysics Source Code Library, November 2000.

\bibitem{Ellerbroek2001}
B.~L. {Ellerbroek} and F.~{Rigaut}.
\newblock {Methods for correcting tilt anisoplanatism in laser-guide-star-based
  multiconjugate adaptive optics}.
\newblock {\em Journal of the Optical Society of America A}, 18:2539--2547,
  October 2001.

\bibitem{Falomo2008}
R.~{Falomo}, A.~{Treves}, J.~K. {Kotilainen}, R.~{Scarpa}, and M.~{Uslenghi}.
\newblock {Near-Infrared Adaptive Optics Imaging of High-Redshift Quasars}.
\newblock {\em \apj}, 673:694--702, February 2008.

\bibitem{Flicker2008}
R~{Flicker}.
\newblock {PSF reconstruction for Keck AO}.
\newblock Technical report, 2008.

\bibitem{Flicker2003}
R.~C. {Flicker}, F.~J. {Rigaut}, and B.~L. {Ellerbroek}.
\newblock {Tilt anisoplanatism in laser-guide-star-based multiconjugate
  adaptive optics. Reconstruction of the long exposure point spread function
  from control loop data}.
\newblock {\em \aap}, 400:1199--1207, March 2003.

\bibitem{Foy1985}
R.~{Foy} and A.~{Labeyrie}.
\newblock {Feasibility of adaptive telescope with laser probe}.
\newblock {\em \aap}, 152:L29--L31, November 1985.

\bibitem{Fried1982}
D.~L. {Fried}.
\newblock {Anisoplanatism in adaptive optics}.
\newblock {\em Journal of the Optical Society of America (1917-1983)}, 72:52,
  January 1982.

\bibitem{Fried1996}
D.~L. {Fried}.
\newblock {Artificial guide star tilt-anisoplanatism: its magnitude and
  amelioration}.
\newblock In M.~{Cullum}, editor, {\em European Southern Observatory Conference
  and Workshop Proceedings}, volume~54 of {\em European Southern Observatory
  Conference and Workshop Proceedings}, page 363, 1996.

\bibitem{Fritz2010}
T.~{Fritz}, S.~{Gillessen}, S.~{Trippe}, T.~{Ott}, H.~{Bartko}, O.~{Pfuhl},
  K.~{Dodds-Eden}, R.~{Davies}, F.~{Eisenhauer}, and R.~{Genzel}.
\newblock {What is limiting near-infrared astrometry in the Galactic Centre?}
\newblock {\em \mnras}, 401:1177--1188, January 2010.

\bibitem{Fusco2000}
T.~{Fusco}, J.-M. {Conan}, L.~M. {Mugnier}, V.~{Michau}, and G.~{Rousset}.
\newblock {Characterization of adaptive optics point spread function for
  anisoplanatic imaging. Application to stellar field deconvolution}.
\newblock {\em \aap}, 142:149--156, February 2000.

\bibitem{Gendron2006}
E.~{Gendron}, Y.~{Cl{\'e}net}, T.~{Fusco}, and G.~{Rousset}.
\newblock {New algorithms for adaptive optics point-spread function
  reconstruction}.
\newblock {\em \aap}, 457:359--363, October 2006.

\bibitem{Ghez2008}
A.~M. {Ghez}, S.~{Salim}, N.~N. {Weinberg}, J.~R. {Lu}, T.~{Do}, J.~K. {Dunn},
  K.~{Matthews}, M.~R. {Morris}, S.~{Yelda}, E.~E. {Becklin}, T.~{Kremenek},
  M.~{Milosavljevic}, and J.~{Naiman}.
\newblock {Measuring Distance and Properties of the Milky Way's Central
  Supermassive Black Hole with Stellar Orbits}.
\newblock {\em \apj}, 689:1044--1062, December 2008.

\bibitem{Gilles2012}
Luc {Gilles}, C~{Correia}, J.P {Véran}, L.~{Wang}, and B~{Ellerbroek}.
\newblock {Simulation model based approach for long exposure atmospheric point
  spread function reconstruction for laser guide star multiconjugate adaptive
  optics}.
\newblock {\em OSA}, 2012.

\bibitem{Guesalaga2016}
A.~{Guesalaga}, B.~{Neichel}, C.~{Correia}, T.~{Butterley}, J.~{Osborn},
  E.~{Masciadri}, T.~{Fusco}, and J.-F. {Sauvage}.
\newblock {Online estimation of atmospheric turbulence parameters and
  outer-scale profiling}.
\newblock In {\em Adaptive Optics Systems V}, volume 9909 of {\em \procspie},
  page 99093C, July 2016.

\bibitem{Jolissaint2010}
L.~{Jolissaint}.
\newblock {Synthetic modeling of astronomical closed loop adaptive optics}.
\newblock {\em Journal of the European Optical Society - Rapid publications, 5,
  10055}, 5, November 2010.

\bibitem{Jolissaint2015}
L.~{Jolissaint}, S.~{Ragland}, and P.~{Wizinowich}.
\newblock {Adaptive Optics Point Spread Function Reconstruction at W. M. Keck
  Observatory in Laser Natural Guide Star Modes : Final Developments}.
\newblock In {\em Adaptive Optics for Extremely Large Telescopes IV (AO4ELT4)},
  page E93, October 2015.

\bibitem{King1983}
I.~R. {King}.
\newblock {Accuracy of measurement of star images on a pixel array}.
\newblock {\em \pasp}, 95:163--168, February 1983.

\bibitem{Lamb2016}
M.~{Lamb}, C.~{Correia}, J.-F. {Sauvage}, D.~{Andersen}, and J.-P. {V{\'e}ran}.
\newblock {Exploring the operational effects of phase diversity for the
  calibration of non-common path errors on NFIRAOS}.
\newblock In {\em Adaptive Optics Systems V}, volume 9909 of {\em \procspie},
  page 99096E, July 2016.

\bibitem{Lu2013}
J.~R. {Lu}, T.~{Do}, A.~M. {Ghez}, M.~R. {Morris}, S.~{Yelda}, and
  K.~{Matthews}.
\newblock {Stellar Populations in the Central 0.5 pc of the Galaxy. II. The
  Initial Mass Function}.
\newblock {\em \apj}, 764:155, February 2013.

\bibitem{Martin2016JATIS}
O.~A. {Martin}, C.~M. {Correia}, E.~{Gendron}, G.~{Rousset}, D.~{Gratadour},
  F.~{Vidal}, T.~J. {Morris}, A.~G. {Basden}, R.~M. {Myers}, B.~{Neichel}, and
  T.~{Fusco}.
\newblock {Point spread function reconstruction validated using on-sky CANARY
  data in multiobject adaptive optics mode}.
\newblock {\em Journal of Astronomical Telescopes, Instruments, and Systems},
  2(4):048001, October 2016.

\bibitem{Martin2016L3S}
O.~A. {Martin}, C.~M {Correia}, E~{Gendron}, G~{Rousset}, F~{Vidal}, T.~J.
  {Morris}, A.~G. {Basden}, R.~M. {Myers}, Y.~H. {Ono}, B.~{Neichel}, and
  T.~{Fusco}.
\newblock {William Herschel Telescope site characterization using the MOAO
  pathfinder CANARY on-sky data}.
\newblock In {\em {Adaptive Optics V from proc. SPIE}}, 2016.

\bibitem{Masciadri2017}
E.~{Masciadri}, F.~{Lascaux}, A.~{Turchi}, and L.~{Fini}.
\newblock {Optical turbulence forecast: ready for an operational application}.
\newblock {\em \mnras}, 466:520--539, April 2017.

\bibitem{Mieda2018}
E~{Mieda}, J.P. {V\'eran}, M~{Rosensteiner}, P~{Turri}, D~{Andersen},
  G~{Herriot}, O~{Lardiere}, and P~{Spano}.
\newblock {Multi-Conjugate Adaptive Optics Simulator for Thirty Meter
  Telescope: Design, Implementation, and Results}.
\newblock {\em Journal of Astronomical Telescopes, Instruments, and Systems},
  2018.

\bibitem{Molodij1997}
G.~{Molodij} and G.~{Rousset}.
\newblock {Angular correlation of Zernike polynomials for a laser guide star in
  adaptive optics.}
\newblock {\em Journal of the Optical Society of America A}, 14:1949--1966,
  August 1997.

\bibitem{Ono2017}
Y.~H. {Ono}, C.~M. {Correia}, D.~R. {Andersen}, O.~{Lardi{\`e}re}, S.~{Oya},
  M.~{Akiyama}, K.~{Jackson}, and C.~{Bradley}.
\newblock {Statistics of turbulence parameters at Maunakea using the multiple
  wavefront sensor data of RAVEN}.
\newblock {\em \mnras}, 465:4931--4941, March 2017.

\bibitem{Osborn2015}
J.~{Osborn}.
\newblock {Scintillation correction for astronomical photometry on large and
  extremely large telescopes with tomographic atmospheric reconstruction}.
\newblock {\em \mnras}, 446:1305--1311, January 2015.

\bibitem{Osborn2013}
J.~{Osborn}, R.~{Wilson}, H.~{Shepherd}, T.~{Butterley}, V.~{Dhillon}, and
  R.~{Avila}.
\newblock {Stereo SCIDAR: Profiling atmospheric optical turbulence with
  improved altitude resolution}.
\newblock In S.~{Esposito} and L.~{Fini}, editors, {\em Proceedings of the
  Third AO4ELT Conference}, December 2013.

\bibitem{Rigaut1998}
F.~J. {Rigaut}, J.-P. {V\'eran}, and O.~{Lai}.
\newblock {\em {Analytical model for Shack-Hartmann-based adaptive optics
  systems}}, volume 3353 of {\em \procspie}, pages 1038--1048.
\newblock September 1998.

\bibitem{Robert2010}
C.~{Robert}, J.-M. {Conan}, D.~{Gratadour}, L.~{Schreiber}, and T.~{Fusco}.
\newblock {Tomographic wavefront error using multi-LGS constellation sensed
  with Shack-Hartmann wavefront sensors}.
\newblock {\em Journal of the Optical Society of America A}, 27(27):A201,
  September 2010.

\bibitem{Roddier1981}
F.~{Roddier}.
\newblock {The effects of atmospheric turbulence in optical astronomy}.
\newblock {\em Progress in optics.~Volume 19.~Amsterdam, North-Holland
  Publishing Co., 1981, p.~281-376.}, 19:281--376, 1981.

\bibitem{Rosensteiner2016}
M.~{Rosensteiner}, P.~{Turri}, E.~{Mieda}, J.-P. {V{\'e}ran}, D.~R. {Andersen},
  and G.~{Herriot}.
\newblock {On the verification of NFIRAOS algorithms and performance on the
  HeNOS bench}.
\newblock In {\em Adaptive Optics Systems V}, volume 9909 of {\em \procspie},
  page 990949, July 2016.

\bibitem{Sarazin2013}
M.~{Sarazin}, M.~{Le Louarn}, J.~{Ascenso}, G.~{Lombardi}, and J.~{Navarrete}.
\newblock {Defining reference turbulence profiles for E-ELT AO performance
  simulations}.
\newblock In S.~{Esposito} and L.~{Fini}, editors, {\em Proceedings of the
  Third AO4ELT Conference}, page~89, December 2013.

\bibitem{Sasiela1994}
R.~J. {Sasiela}.
\newblock {Wave-front correction using one or more synthetic beacons.}
\newblock {\em Journal of the Optical Society of America A}, 11:379--393,
  January 1994.

\bibitem{Saxenhuber2017}
D.~{Saxenhuber}, G.~{Auzinger}, M.~L. {Louarn}, and T.~{Helin}.
\newblock {Comparison of methods for the reduction of reconstructed layers in
  atmospheric tomography}.
\newblock {\em \ao}, 56:2621, April 2017.

\bibitem{Shodel2010}
R.~{Sch{\"o}del}.
\newblock {Accurate photometry with adaptive optics in the presence of
  anisoplanatic effects with a sparsely sampled PSF. The Galactic center as an
  example of a challenging target for accurate AO photometry}.
\newblock {\em \aap}, 509:A58, January 2010.

\bibitem{Schramm2013}
M.~{Schramm} and J.~D. {Silverman}.
\newblock {The Black Hole-Bulge Mass Relation of Active Galactic Nuclei in the
  Extended Chandra Deep Field-South Survey}.
\newblock {\em \apj}, 767:13, April 2013.

\bibitem{Sheehy2006}
C.~D. {Sheehy}, N.~{McCrady}, and J.~R. {Graham}.
\newblock {Constraining the Adaptive Optics Point-Spread Function in Crowded
  Fields: Measuring Photometric Aperture Corrections}.
\newblock {\em \apj}, 647:1517--1530, August 2006.

\bibitem{Stetson1987}
P.~B. {Stetson}.
\newblock {DAOPHOT - A computer program for crowded-field stellar photometry}.
\newblock {\em \pasp}, 99:191--222, March 1987.

\bibitem{Tokovinin2007}
A.~{Tokovinin} and V.~{Kornilov}.
\newblock {Accurate seeing measurements with MASS and DIMM}.
\newblock {\em \mnras}, 381:1179--1189, November 2007.

\bibitem{Turri2017}
P.~{Turri}, A.~W. {McConnachie}, P.~B. {Stetson}, G.~{Fiorentino}, D.~R.
  {Andersen}, G.~{Bono}, D.~{Massari}, and J.-P. {V{\'e}ran}.
\newblock {Optimal Stellar Photometry for Multi-conjugate Adaptive Optics
  Systems Using Science-based Metrics}.
\newblock {\em The Astronomical Journal}, 153:199, April 2017.

\bibitem{Tyler1994}
G.~A. {Tyler}.
\newblock {Merging: a new method for tomography through random media}.
\newblock {\em Journal of the Optical Society of America A}, 11:409--424,
  January 1994.

\bibitem{VanDam2006}
M.~A. {van Dam}, R.~J. {Sasiela}, A.~H. {Bouchez}, D.~{Le Mignant}, R.~D.
  {Campbell}, J.~C.~Y. {Chin}, S.~K. {Hartman}, E.~M. {Johansson}, R.~E.
  {Lafon}, P.~J. {Stomski}, Jr., D.~M. {Summers}, and P.~L. {Wizinowich}.
\newblock {Angular anisoplanatism in laser guide star adaptive optics}.
\newblock In {\em Society of Photo-Optical Instrumentation Engineers (SPIE)
  Conference Series}, volume 6272 of {\em \procspie}, page 627231, June 2006.

\bibitem{Veran1997}
J.-P. {Veran}, F.~{Rigaut}, H.~{Maitre}, and D.~{Rouan}.
\newblock {Estimation of the adaptive optics long-exposure point-spread
  function using control loop data.}
\newblock {\em Journal of the Optical Society of America A}, 14:3057--3069,
  November 1997.

\bibitem{Wilson2002}
R.~W. {Wilson}.
\newblock {SLODAR: measuring optical turbulence altitude with a Shack-Hartmann
  wavefront sensor}.
\newblock {\em \mnras}, 337:103--108, November 2002.

\bibitem{Winick1988}
K.~A. {Winick} and D.~V.~L. {Marquis}.
\newblock {Stellar scintillation technique for the measurement of tilt
  anisoplanatism}.
\newblock {\em Journal of the Optical Society of America A}, 5:1929--1936,
  November 1988.

\bibitem{Yelda2010}
S.~{Yelda}, J.~R. {Lu}, A.~M. {Ghez}, W.~{Clarkson}, J.~{Anderson}, T.~{Do},
  and K.~{Matthews}.
\newblock {Improving Galactic Center Astrometry by Reducing the Effects of
  Geometric Distortion}.
\newblock {\em \apj}, 725:331--352, December 2010.

\end{thebibliography}

\end{document}